\documentclass[12pt]{article}

\usepackage[margin=1in]{geometry}
\usepackage[T1]{fontenc}
\usepackage[utf8]{inputenc}
\usepackage{amsmath,amssymb,amsthm,mathtools}
\usepackage{booktabs}
\usepackage{graphicx}
\usepackage{placeins}
\usepackage{authblk}
\usepackage[round,authoryear]{natbib}
\usepackage[colorlinks=true,linkcolor=blue,citecolor=blue,urlcolor=blue]{hyperref}

\theoremstyle{plain}
\newtheorem{theorem}{Theorem}[section]

\newtheorem{corollary}[theorem]{Corollary}
\newtheorem{lemma}[theorem]{Lemma}
\theoremstyle{remark}
\newtheorem{remark}[theorem]{Remark}
\theoremstyle{plain}

\DeclareMathOperator{\sech}{sech}
\newcommand{\R}{\mathbb R}
\newcommand{\E}{\mathbb E}
\newcommand{\Prob}{\mathbb P}
\newcommand{\dd}{\,\mathrm d}

\title{Diffusion limit of the 2PL model}
\author{Sourav Majumdar\thanks{souravm@iitk.ac.in}\\Nivedita Bhaktha\thanks{nbhaktha@iitk.ac.in}}
\affil{Department of Management Sciences, Indian Instititute of Technology Kanpur, India}
\date{}

\begin{document}
\maketitle

\begin{abstract}

Item response theory specifies accuracy probabilities without a response process or timing mechanism. We show that thresholding at zero the stationary state of a diffusion with a logistic invariant distribution recovers the two-parameter logistic (2PL) probability for every starting value. A two-boundary response rule gives a logistic diffusion IRT model for accuracy and response time. Through the connection between the diffusion's forward equation and the P\"oschl--Teller potential, we derive explicit exit probabilities, Laplace transforms, densities and means of first passage times, and the joint density of decision time and response. Item difficulty and discrimination have different roles from evidence thresholds and time scales. The model predicts how accuracy varies with response time above and below item difficulty. In simulations, response times improved ability estimates, and lowering the evidence threshold reduced their accuracy more than halving test length. In PISA 2018 mathematics data, the model identified different time demands for similarly difficult items, predicted withheld answers nearly as accurately as the hierarchical model, and identified where its account of accuracy and response time failed.

\end{abstract}


\section{Introduction}

Item response theory (IRT) is the standard framework for modelling responses to test items, questionnaires, and related instruments, and it underlies scoring in large-scale educational assessments, licensure testing, standardized testing, and patient-reported outcome measurement. In a typical binary IRT model, the response to a test item can be correct or incorrect, and the probability of a correct response is modelled as a function of a person dependent ability parameter and item parameters. The Rasch model and the two parameter logistic (2PL) model are the most commonly used IRT models, see the reviews by \citet{cai2016item} and \citet{chen2025item}. In the 2PL model,
\begin{equation}\label{eq:intro-2pl}
        \Prob(Y_{ip}=1)
        =\frac{1}{1+\exp\{-a_i(\theta_p-\beta_i)\}},
\end{equation}
where \(\theta_p\) is the ability of person \(p\), \(\beta_i\) is the difficulty of item \(i\), and \(a_i\) is the discrimination of item \(i\).

The 2PL model specifies the probability of a correct response, but it does not describe how a response develops over time, on its own. This limitation has become more consequential as assessments have moved to computer-based administration. Large-scale international assessments such as PISA now record time taken on every item and the response times are being used to improve the precision of ability estimates, to detect disengaged (careless) responding, and to study the cognitive process of test taking \citep{van2009conceptual, de2019overview}. To assign a probability to both accuracy and response time requires a mechanism that determines which response occurs and when it occurs. 

Joint models of response accuracy and response time specify both the performance and the time taken to produce a response in a test. These approaches differ in how they associate response accuracy to response time and interpret the latent variables used to model them \citep{de2019overview}. In the hierarchical framework \citep{van2007hierarchical} an IRT model of accuracy and a model for response time are linked through a joint distribution of person ability and speed. \citet{van2006lognormal} modelled response time using a lognormal distribution with person speed and item time intensity parameters. Box--Cox transformations allow for different shapes of response time distributions \citep{kleinentink2009boxcox}. \citet{wang2013semiparametric} used a Cox proportional hazards model with an unspecified baseline hazard for each item. The bivariate generalized linear IRT framework of \citet{molenaar2015bglirt} links models for responses and response times through linear, nonlinear, or interaction terms. \citet{liu2024semiparametric} used semiparametric factor analysis to study response accuracy and time. \citet{kwon2026copula} used copulas to model dependence between latent ability and speed.  These models employ a standard IRT model for accuracy and describe response time using a different statistical model, but they do not specify the process by which a response is produced.

A consequence is that accuracy and response time can remain associated after accounting for person and item characteristics and this association must be modelled separately. \citet{bolsinova2017conditional} modelled conditional dependence by allowing standardized residual log response time to affect the intercept and discrimination of the item response function. Response moderation models allow the association between residual log response time and accuracy to vary across items, persons, or both \citep{bolsinova2017moderation}, and further extensions accommodate nonlinear conditional relationships \citep{bolsinova2018nonlinear}. \citet{bolsinova2019differences} approached the same issue through response time models whose parameters can differ between correct and incorrect responses. In these approaches, the direction and size of the conditional dependence are estimated rather than implied by the model for the response process. 

Diffusion and race models describe accuracy and time through a stochastic response process. In a diffusion model, accumulated evidence changes until it reaches one of two decision boundaries. The boundary reached determines the response, and the time taken to reach it is the decision time \citep{ratcliff1978theory,ratcliff2016diffusion}. Race models assign a finishing time to each competing response process. The first process to finish determines the response and its decision time. Examples include the Birnbaum--Saunders race model of \citet{ranger2015race} and the lognormal race model of \citet{rouder2015lognormal}. \citet{ranger2014accumulator} used proportional hazards to model competing processes of knowledge acquisition and discontinuation. \citet{jin2023accumulator} extended an accumulator model with latent person-item interactions. Conditional dependence has also motivated extensions of diffusion IRT models. \citet{kang2022modeling} added starting-point variability to a diffusion IRT model with drift-rate variability, allowing it to account for different conditional associations between accuracy and response time. \citet{kang2023latent} used distances between persons and items in a latent space to model interactions in drift rate beyond the main person and item effects.

Process models described by the diffusion and race models generally yield response probabilities that differ from the 2PL model. \citet{tuerlinckx2005two} derived the 2PL response function from diffusion and race models and provided a stochastic process based interpretation of item discrimination. In their diffusion construction, the process starts midway between two boundaries, and the 2PL probability is the probability of reaching the upper boundary before the lower boundary. With the diffusion scale fixed, discrimination corresponds to the distance between the boundaries. \citet{maris2012speedaccuracy} derived a joint response model from a scoring rule that awards the time remaining before an item deadline for a correct response and deducts it for an incorrect response. Its marginal probability of a correct response has the 2PL form, with the item deadline determining discrimination. \citet{tijmstra2021speed} subsequently incorporated individual differences in speed into a discretized version of this model. In both constructions, the 2PL depends on a feature of the decision rule, the boundary separation or the dealine, and discrimination is interpreted through that feature. 

The present work demonstrates that the 2PL probability arises as an intrinsic property of a specific evidence accumulation process itself, rather than from the decision rule. We study a diffusion whose drift moves the accumulated evidence towards $\mu_{ip} = a_i(\theta_p -\beta _i)$ and whose stationary distribution is logistic with $\mu_{ip}$ as its location. The 2PL probability is the probability that a diffusion with a logistic stationary distribution lies above zero at stationarity. As the process approaches stationarity, this probability converges to the 2PL probability irrespective of the starting value; and ability, difficulty and discrimination parameters retain their usual interpretation. The 2PL response probability thus emerges as a long-run property of the proposed evidence accumulation process. 

To model response time, we study the first passage properties of the logistic diffusion. A response occurs when the process first reaches one of two boundaries placed symmetrically about the starting value. The resulting response probabilities generally differ from the usual 2PL probabilities, allowing us to study response accuracy under a finite-boundary decision rule. Each item has an evidence threshold, which determines how  much evidence is needed before a response is produced. Therefore, two items can have the same probability of a correct response for a given respondent even if the response time for one of the items is longer than the other.

The contributions of this paper are twofold. The first is conceptual, establishing a diffusion interpretation of the 2PL model in which the 2PL probability is the stationary probability of a correct response. The second is methodological. We develop the \emph{logistic diffusion IRT model} for response accuracy and response time, named after the logistic stationary distribution of the evidence accumulation process. We exploit the connection between the diffusion's forward equation and the P\"oschl--Teller potential \citep{brics2013solve} to derive explicit expressions for boundary-hitting probabilities and response time distributions. The model also yields testable predictions in which early responses are less accurate than later ones for a fixed ability above an item difficulty, while the reverse holds below item difficulty. We evaluate the resulting model in three simulation studies, which examine the parameter recovery, the contribution of response times to ability estimation, and the sensitivity of ability estimates to departure from the model, and in an application to PISA 2018 mathematics data. 

The paper is organized as follows. In Section \ref{sec:logistic-diffusion}, we present the logistic diffusion model and its 2PL stationary distribution. In Section \ref{sec:joint-model}, we present the joint reponse time and accuracy model based on the logistic diffusion process. We present the results of a simulation study and evaluation of the joint model on PISA 2018 data in Section \ref{sec:evaluation}. Finally in Section \ref{sec:discussion} we conclude the article.

\section{A diffusion interpretation of the 2PL model}\label{sec:logistic-diffusion}

The 2PL model provides the probability that a person of a given ability answers an item correctly. We score the response of person $p$ to item $i$ as correct if the accumulated evidence $Z_t$ is above zero at the time of observation, and as incorrect otherwise. We show that, at stationarity, the probability that $Z_t$ is above zero equals the 2PL probability of a correct response. We model the evidence accumulation process by a diffusion $(Z_t)_{t\ge0}$ satisfying
\begin{equation}\label{eq:main-diffusion}
    dZ_t
    =-\frac{\sigma^2}{2}
      \tanh\left(\frac{Z_t-\mu}{2}\right)dt
      +\sigma dW_t,
    \qquad \mu\in\mathbb R,\quad \sigma>0,
\end{equation}
where $(W_t)_{t\geq 0}$ is standard Brownian motion. Over a short interval, evidence tends to increase if it starts below $\mu$ and to decrease if it starts above $\mu$. The drift is zero at $\mu$. The Brownian motion as the noise term allows the evidence process on any one path to move up or down.

We compare values of $\sigma$ for processes with the same starting value $Z_0=x$. For a smooth function $f$, its infinitesimal generator $(\mathcal L f)(x)$ is the rate at which the expectation of $f(Z_t)$ changes at $t=0$ \citep[Section 7.3]{oksendal2003}. Choosing $f(z)=z$ recovers the drift at the starting state. For the proposed diffusion, the generator takes the form
\begin{equation}\label{eq:L-generator}
    (\mathcal L f)(x)
    = \frac{\sigma^2}{2}
      \left[
        \frac{\mathrm{d}^2 f(x)}{\mathrm{d} x^2}
        - \tanh\left(\frac{x-\mu}{2}\right)\frac{\mathrm{d} f(x)}{\mathrm{d} x}
      \right].
\end{equation}
Doubling $\sigma$ multiplies the generator by four. With the starting value held fixed, observing this faster process at time $t$ is equivalent in distribution to observing the original process at $4t$. In this sense $\sigma$ controls the rate of convergence to stationarity.

The diffusion term in $Z_t$ tends to spread the values of the evidence, and the drift term tends to bring them back towards $\mu$. At stationarity the evidence on a given path still changes, but observing the process later gives the same distribution. The resulting distribution in \eqref{eq:main-diffusion} is logistic with location $\mu$ and unit scale,
\begin{equation}\label{eq:logistic-density}
    \pi_\mu(x)=\frac14\sech^2\left(\frac{x-\mu}{2}\right).
\end{equation}

A response is scored as correct when the accumulated evidence is above zero. Under this density, the probability above zero is $1/(1+\exp(-\mu))$. When the location is the item's discrimination times the difference between person ability and item difficulty, $\mu=a_i(\theta_p-\beta_i)$, this probability is exactly the 2PL response probability.

\begin{theorem}[Invariant law and logistic threshold probability]\label{thm:invariant}
\eqref{eq:main-diffusion} admits a unique non-explosive strong solution for each \(Z_0=x\in\R\). This process is positive recurrent with invariant density \(\pi_\mu\). Moreover,
\begin{equation}\label{eq:logistic-tail}
        \lim_{t\to\infty}\Prob_x(Z_t>0)
        =\int_0^\infty \pi_\mu(z)\dd z
        =\frac{1}{1+\exp(-\mu)}.
\end{equation}
\end{theorem}

The proof is given in Appendix \ref{app:proof-invariant}. We can see that Theorem~\ref{thm:invariant} has two parts. The first ensures that the evidence accumulation process is well defined for every starting value and does not diverge in finite time. The second describes its long run behavior. The distribution of the accumulated evidence settles to the logistic density $\pi_\mu$, and the probability that the evidence is positive converges to the logistic function of $\mu$. The limit does not depend on the starting value x. Whatever evidence a respondent holds when an item is presented, for example, an initial impression formed on reading the item, its influence on the probability of a positive value vanishes as the process approaches stationarity. The noise parameter $\sigma$ also does not appear in the limit. The diffusion generator determines how quickly the process approaches stationarity, but not the stationary probability itself.

\begin{corollary}[Stationary representation of the 2PL model]\label{cor:2pl}
Set
\begin{equation}\label{eq:irt-mu}
        \mu_{ip}=a_i(\theta_p-\beta_i).
\end{equation}
Let \(Z^{ip}\) solve \eqref{eq:main-diffusion} with \(\mu=\mu_{ip}\), and define
\[
        Y_t^{ip}=\mathbf 1_{\{Z_t^{ip}>0\}}.
\]
Then
\begin{equation}\label{eq:2pl-link}
        \lim_{t\to\infty}\Prob(Y_t^{ip}=1)
        =\frac{1}{1+\exp\{-a_i(\theta_p-\beta_i)\}}.
\end{equation}
\end{corollary}

The proof is given in Appendix \ref{app:proof-2pl}. We can see that Corollary~\ref{cor:2pl} demonstrates that the 2PL model is the stationary response model of the evidence accumulation process when $\mu_{ip} = a_i(\theta_p - \beta _i)$. Ability and difficulty arise in the difference, so the process for a given person and item depends on how far, and in which direction, ability lies from difficulty. Discrimination converts this difference into the location of the stationary distribution. For a highly discriminating item, a unit difference in ability moves the stationary distribution further from zero, and the probability of a positive value changes more sharply near the item difficulty. All three parameters therefore retain their usual interpretation. This construction differs from the diffusion process interpretation of \citet{tuerlinckx2005two} in how the 2PL probability arises. There, it is the probability of reaching the upper boundary first from a starting point midway between them, and discrimination corresponds to the distance between the boundaries. Whereas in Corollary~\ref{cor:2pl} no boundary is involved. The 2PL probability is a property of the stationary distribution alone. A boundary enters only when response time is modelled in Section \ref{sec:joint-model}, where it is a parameter separate from discrimination. 

In the basic Ratcliff model, the quality of the information available to the respondent determines a drift rate that stays constant during the decision \citep{ratcliff2016diffusion}. In our diffusion process the drift term depends on the distance and direction from the current state to $\mu_{ip}$. When ability exceeds item difficulty, the state tends towards a positive value. If the evidence happens to be negative when observed, the response is incorrect. Section \ref{sec:joint-model} uses the boundaries in Figure \ref{fig:response-paths} to score a response. For a given item and ability, Figure \ref{fig:accuracy-differences} shows how the resulting probability of success differs from the stationary 2PL probability. The drift in Equation~\ref{eq:main-diffusion} is bounded, and its magnitude never exceeds $\sigma^2/2$. Far from $\mu_{ip}$, the process therefore behaves approximately like Brownian motion with a constant drift of $\pm \sigma^2/2$ towards $\mu_{ip}$, as in the basic Ratcliff model. Close to $\mu_{ip}$, the drift is approximately proportional to the distance from $\mu_{ip}$, so evidence is drawn back towards the value determined by ability, difficulty and discrimination with a force that weakens as it approaches that value.

\begin{remark}[Connection with the P\"oschl-Teller potential]\label{rem:pt}
For each possible answer, we need the probability of reaching its boundary first and doing so by time $t$. A change of variables puts this problem in a form with a known solution in physics. Centring and rescaling the state, then factoring the transition density, removes the first spatial derivative from the equation for $R$. Let
\[
        U_t=\frac{Z_t-\mu}{2},
        \qquad D=\frac{\sigma^2}{8},
\]
and let \(q(t,u\mid u_0)\) be the transition density of \(U\). If we write
\[
        q(t,u\mid u_0)=\frac{\sech u}{\sech u_0}R(t,u\mid u_0),
\]
then the forward equation for \(q\) becomes
\[
        \frac{\partial R}{\partial t}
        =D\frac{\partial^2 R}{\partial u^2}-D\{1-2\sech^2u\}R.
\]
The spatial part of this equation also appears in quantum mechanics for the P\"oschl--Teller potential \citep{brics2013solve}. We use the spatial solutions for this potential to obtain the distribution of the waiting time until the first of the two response boundaries is reached.
\end{remark}

Remark~\ref{rem:pt} identifies the reason the first-passage problems of Section~\ref{sec:joint-model} are tractable. After the change of variables, the transformed density satisfies an equation whose spatial part is the P\"oschl--Teller problem, for which the eigenfunctions are known explicitly \citep{brics2013solve}. As a result, the one-threshold first passage time densities are available in closed form (Theorem~\ref{thm:fpt-density}), and the two-boundary joint density is an eigenfunction expansion with explicitly known terms (Theorem~\ref{thm:joint-exit}). Response time distributions can therefore be evaluated directly rather than by numerical methods.

\section{Modelling response accuracy and response time}\label{sec:joint-model}

Following \citet{ratcliff2016diffusion}, we assume that respondents answer once they have accumulated enough evidence for one of the two outcomes. From a starting value of zero, evidence accumulates in favour of a correct or an incorrect answer. The two response thresholds are the same distance $A$ from zero. Evidence accumulation ends with a correct response on reaching $+A$ or an incorrect response on reaching $-A$, whichever occurs first. Figure \ref{fig:response-paths} shows how long the respondent spends accumulating evidence before answering. This is the decision time.

\subsection{Waiting time for one response threshold}\label{sec:one-sided}

For this calculation, evidence continues to accumulate until it reaches the threshold of interest. A respondent of higher ability takes less time on average to reach $+A$ and more time on average to reach $-A$. For instance, a path can pass through $-A$ before reaching $+A$. It counts towards the waiting time for $+A$ in this calculation, even though a respondent facing both boundaries would have answered incorrectly at $-A$. For a path starting at $Z_0=x$, the time to reach $c\in\mathbb R$ is
\begin{equation}\label{eq:tau-c}
    \tau_c=\inf\{t\ge0:Z_t=c\}.
\end{equation}
We call $\tau_c$ the first passage time.

The calculation uses the Laplace transform of this waiting time,
\[
    \phi_\lambda(x;c)=\E_x[e^{-\lambda\tau_c}],
    \qquad \lambda\ge0,
\]
where $\E_x$ is expectation given $Z_0=x$. Short waiting times receive more weight than long ones for $\lambda>0$. The inverse transform yields the density of $\tau_c$ for a threshold on either side of $x$. 

\begin{theorem}[Laplace transform of the one-threshold hitting time]\label{thm:laplace-fpt}
Let
\begin{equation}\label{eq:w-defs}
        w=\frac{x-\mu}{2},
        \qquad
        w_c=\frac{c-\mu}{2},
        \qquad
        \nu=\sqrt{1+\frac{8\lambda}{\sigma^2}}.
\end{equation}
If \(x<c\) and \(\lambda>0\), then
\begin{equation}\label{eq:up-laplace}
        \phi_\lambda(x;c)
        =\frac{\cosh w}{\cosh w_c}
        e^{-\nu(w_c-w)}
        \frac{\tanh w-\nu}{\tanh w_c-\nu}.
\end{equation}
If \(x>c\) and \(\lambda>0\), then
\begin{equation}\label{eq:down-laplace}
        \phi_\lambda(x;c)
        =\frac{\cosh w}{\cosh w_c}
        e^{-\nu(w-w_c)}
        \frac{\tanh w+\nu}{\tanh w_c+\nu}.
\end{equation}

\end{theorem}

The proof is given in Appendix \ref{app:proof-laplace-fpt}.

The first passage time density uses the constant
\begin{equation}\label{eq:gamma}
        \delta=\frac{8}{\sigma^2}.
\end{equation}
The inversion identities used to obtain the density are collected in Appendix \ref{app:laplace-identities}. The Laplace transform is used to obtain the distribution of the waiting time to a single threshold. Its derivative at $\lambda = 0$ gives the mean waiting time (Theorem~\ref{thm:mean-fpt}), and its inversion gives the density (Theorem~\ref{thm:fpt-density}). The transform depends on the starting value, the threshold and $\mu$ only through $w$ and $w_c$, which measure the starting value and the threshold relative to the location of the stationary distribution. The waiting time to a threshold therefore depends on where the threshold and the starting value lie relative to $\mu_{ip} = a_i(\theta_p-\beta_i)$, not on ability and difficulty separately. 

\begin{theorem}[First passage time density]\label{thm:fpt-density}
Assume \(x\ne c\). The law of \(\tau_c\) has a density on \((0,\infty)\).

If \(x<c\), define
\begin{equation}\label{eq:ABC-up}
        d=(w_c-w)\sqrt\delta,
        \qquad
        B=\frac{\tanh w_c}{\sqrt\delta},
        \qquad
        C_+=(\tanh w_c-\tanh w)\sqrt\delta.
\end{equation}
Then, for \(t>0\),
\begin{align}\label{eq:fpt-density-up}
        f_{\tau_c}(t\mid x)
        &=\frac{\cosh w}{\cosh w_c}e^{-t/\delta}
        \Bigg[
        \frac{d}{2\sqrt\pi\,t^{3/2}}
        e^{-d^2/(4t)} \notag\\
        &\qquad
        +\frac{C_+}{\delta}
        \left\{
        \frac{1}{\sqrt{\pi t}}e^{-d^2/(4t)}
        +B e^{-dB+B^2t}
        \operatorname{erfc}\left(\frac{d}{2\sqrt t}-B\sqrt t\right)
        \right\}
        \Bigg].
\end{align}

If \(x>c\), define
\begin{equation}\label{eq:ABC-down}
        d=(w-w_c)\sqrt\delta,
        \qquad
        B=\frac{\tanh w_c}{\sqrt\delta},
        \qquad
        C_-=(\tanh w-\tanh w_c)\sqrt\delta.
\end{equation}
Then, for \(t>0\),
\begin{align}\label{eq:fpt-density-down}
        f_{\tau_c}(t\mid x)
        &=\frac{\cosh w}{\cosh w_c}e^{-t/\delta}
        \Bigg[
        \frac{d}{2\sqrt\pi\,t^{3/2}}
        e^{-d^2/(4t)} \notag\\
        &\qquad
        +\frac{C_-}{\delta}
        \left\{
        \frac{1}{\sqrt{\pi t}}e^{-d^2/(4t)}
        -B e^{dB+B^2t}
        \operatorname{erfc}\left(\frac{d}{2\sqrt t}+B\sqrt t\right)
        \right\}
        \Bigg].
\end{align}
If \(x=c\), then \(\tau_c=0\) almost surely.
\end{theorem}

The proof is given in Appendix \ref{app:proof-fpt-density}. Theorem~\ref{thm:fpt-density} gives the first passage time density in closed form, so it can be evaluated directly and differentiated with respect to the parameters. The factor $e^{-t/\delta}$, with $\delta = 8/\sigma^2$, makes the density decay exponentially at a rate that increases with $\sigma^2$, which is consistent with the role of $\sigma$ as the speed of convergence identified after Equation~\ref{eq:L-generator}. The factor $\cosh w / \cosh w_c$ reflects the position of the starting value and the threshold relative to $\mu_{ip}$.

\begin{theorem}[Mean first passage time]\label{thm:mean-fpt}
The mean first passage time \(m_c(x)=\E_x[\tau_c]\) is
\begin{equation}\label{eq:mfpt}
        m_c(x)=
        \begin{cases}
        \displaystyle
        \frac{2}{\sigma^2}\left[(c-x)+e^{c-\mu}-e^{x-\mu}\right],
        & x\le c,\\[1.25em]
        \displaystyle
        \frac{2}{\sigma^2}\left[(x-c)+e^{-(c-\mu)}-e^{-(x-\mu)}\right],
        & x\ge c.
        \end{cases}
\end{equation}
\end{theorem}

The proof is given in Appendix \ref{app:proof-mean-fpt}.

At a fixed ability level, moving the threshold $c$ farther from the starting point $x$ on the same side increases the mean time to reach it (Theorem \ref{thm:mean-fpt}). With $c$ fixed, higher ability increases $\mu=a_i(\theta_p-\beta_i)$ for $a_i>0$, shifting the drift upwards. The expected time then falls if $c>x$ and rises if $c<x$. Making the item easier has the same effect as raising ability by the same amount. Two further features of Equation~\ref{eq:mfpt} are relevant for response times. First, for a threshold above the starting value, the mean waiting time decreases as $\mu$ increases but approaches $2(c - x)/\sigma^2$ rather than zero, because $e^{c-\mu} - e^{x-\mu}$ tends to zero. However able a respondent is, the mean time to accumulate evidence to an upper threshold cannot fall below this value, because the drift never exceeds $\sigma^2/2$. Second, when $\mu$ is far below the threshold, the mean waiting time grows like $e^{c-\mu}$: reaching a threshold that lies against the drift becomes exponentially slow. Both features follow from the bounded drift in Equation~\ref{eq:main-diffusion} and distinguish the process from a diffusion whose drift grows without bound with ability.

\subsection{Correct and incorrect responses at two boundaries}\label{sec:two-sided}

The single-threshold distribution for the correct boundary $r$ includes paths that crossed the incorrect boundary $\ell<r$ on the way to $r$. Each distribution gives the waiting time to one boundary. A correct response by time $t$ requires both $\tau_r\le t$ and $\tau_r<\tau_\ell$ on the same path. The decision time is
\begin{equation}\label{eq:tau-interval}
        \tau_{\ell,r}=\min(\tau_\ell,\tau_r)=\inf\{t\ge0:Z_t\notin(\ell,r)\}.
\end{equation}
The response indicator is
\[
        Y=\mathbf 1_{\{Z_{\tau_{\ell,r}}=r\}}.
\]

\begin{theorem}[Exit probability]\label{thm:exit-prob}
For \(x\in(\ell,r)\),
\begin{equation}\label{eq:exit-prob}
        p_{\ell,r}(x):=\Prob_x(Y=1)
        =\frac{S(x)-S(\ell)}{S(r)-S(\ell)},
\end{equation}
where
\[
        S(x)=x-\mu+\sinh(x-\mu)
\]
is the scale function in \eqref{eq:scale}.
\end{theorem}

The proof is given in Appendix \ref{app:proof-exit-prob}.

Integrating the upper-boundary density over a time interval gives the probability of a correct response during that interval. The lower-boundary density gives the corresponding probability for an incorrect response. The transformed coordinates and diffusion scale are
\begin{equation}\label{eq:w-bounds}
        u=\frac{z-\mu}{2},
        \qquad
        u_0=\frac{x-\mu}{2},
        \qquad
        u_\ell=\frac{\ell-\mu}{2},
        \qquad
        u_r=\frac{r-\mu}{2},
        \qquad
        D=\frac{\sigma^2}{8}.
\end{equation}

Each term in the joint density corresponds to one distinct Dirichlet eigenvalue \(\lambda_n\) for the interval \((u_\ell,u_r)\). Appendix \ref{app:transition-density} specifies the eigenfunctions \(\Psi_n\) and their squared norms \(N_n\). Theorem~\ref{thm:exit-prob} gives the probability of a correct response under the two-boundary rule. The scale function $S$ transforms the process into one without drift, and the probability of exiting at $r$ is the relative position of the starting value between $\ell$ and $r$ on that scale. Unlike the stationary probability in Corollary~\ref{cor:2pl}, the exit probability depends on the starting value and the boundaries. Under the two-boundary rule, the probability of a correct response therefore reflects the person--item difference, through $\mu_{ip}$, and the decision rule, through $\ell$, $r$, and $x$. Section~\ref{sec:correct-probability} examines what this implies for the item response function.

\begin{theorem}[Joint density of exit time and exit side]\label{thm:joint-exit}
For \(t>0\) and \(x\in(\ell,r)\), the joint densities of \(\tau_{\ell,r}\) and the exit side are
\begin{align}\label{eq:exit-density-upper}
        f_{\tau,Y}(t,1\mid x)
        &=D\frac{\sech u_r}{\sech u_0}
        \sum_{n\ge1}e^{-\lambda_nt}
        \frac{\{-\left.\frac{\mathrm{d} \Psi_n(u)}{\mathrm{d} u}\right|_{u=u_r}\}\Psi_n(u_0)}{N_n},\\[0.5em]
        f_{\tau,Y}(t,0\mid x)
        &=D\frac{\sech u_\ell}{\sech u_0}
        \sum_{n\ge1}e^{-\lambda_nt}
        \frac{\left.\frac{\mathrm{d} \Psi_n(u)}{\mathrm{d} u}\right|_{u=u_\ell}\Psi_n(u_0)}{N_n}.
        \label{eq:exit-density-lower}
\end{align}
Therefore
\begin{equation}\label{eq:exit-density-total}
        f_\tau(t\mid x)
        =f_{\tau,Y}(t,1\mid x)+f_{\tau,Y}(t,0\mid x).
\end{equation}
\end{theorem}

\begin{figure}[!htbp]
\centering
\includegraphics[width=\textwidth]{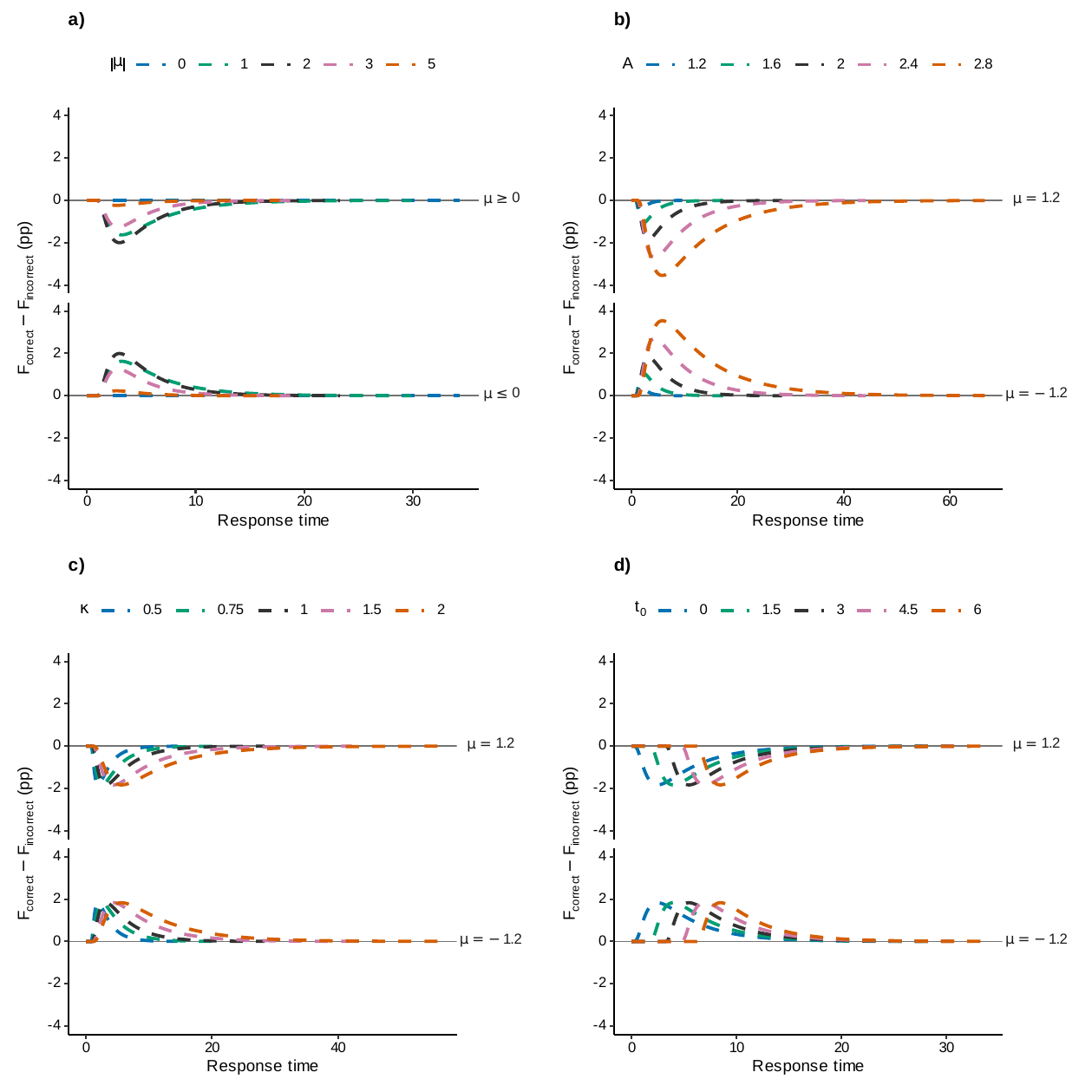}
\caption{Difference in cumulative response time probabilities (correct minus incorrect), in percentage points. Negative values indicate that a greater fraction of incorrect responses have occurred by the indicated time. Panel (a) varies $\mu$. Panels (b)--(d) vary $A$, $\kappa$, and $t_0$, respectively, with $\mu=1.2$ above and $\mu=-1.2$ below. Unless varied, $A=2$, $\kappa=1$, $t_0=0.7$, and $\sigma=1$.}
\label{fig:conditional-rt-differences}
\end{figure}

The proof is given in Appendix \ref{app:proof-joint-exit}. The joint density decomposes the decision time distribution by response. Both components share the exponential factors $e^{-\lambda_n t}$, and for large $t$ the term with the smallest eigenvalue $\lambda_1$ dominates. The right tails of the decision time distributions for correct and incorrect responses therefore decay at the same rate, and differences between them are concentrated at short and intermediate decision times, as Figure~\ref{fig:conditional-rt-differences} shows. In the diffusion with constant drift and a starting point midway between the boundaries, correct and incorrect decision times have the same distribution. Here the drift depends on the current state, so this symmetry does not hold, and the model generates a dependence between accuracy and response time at fixed person and item parameters.

Figure \ref{fig:conditional-rt-differences}a gives a prediction that can be checked by comparing early and later responses to one item among students of the same ability. When that ability exceeds the item's difficulty, evidence can reach the incorrect boundary before the drift draws it towards the correct side. The model therefore predicts a larger proportion of incorrect answers among students who respond by time $t$ than among those who respond later. At an ability equally far below the item's difficulty, the drift acts in the opposite direction, and the model predicts that a larger proportion of the early answers will be correct. In Figure~\ref{fig:conditional-rt-differences}b--d, the evidence threshold, time scale and non-decision time change the size of these differences but not their direction, which is determined by the sign of $\theta_p - \beta_i$. The conditional dependence is thus a consequence of the response process rather than an additional model component, and its direction is a testable prediction.

The empirical findings are mixed. \citet{kang2022modeling} discuss data in which accuracy rose with response time on easy items and fell on difficult items over the shorter response times. At longer response times, accuracy fell on the easy items and rose on the difficult ones. In a low-stakes educational test, \citet{bolsinova2017conditional} found that relatively fast responses tended to be more accurate on easier items and relatively slow responses more accurate on harder items.

\subsection{Person and item parameters in the joint model}\label{sec:application}

Two items can have the same probability of a correct response for a given student even if one takes longer to answer. Items can differ both in how much evidence a respondent needs and in how long it takes to accumulate that evidence. For person \(p\) answering item \(i\), ability, difficulty and discrimination determine the value towards which the evidence tends,
\[
        \mu_{ip}=a_i(\theta_p-\beta_i).
\]
Scoring the sign of the evidence at stationarity gives the 2PL probability (Corollary \ref{cor:2pl}). Under the two-boundary rule, the probability of a correct response also depends on the item's evidence threshold. Its time scale converts the time to reach that threshold into units of observed response time.

\subsubsection{The response rule and observed response time}\label{sec:response-rule}

An observed response occurs when the evidence accumulation process first reaches a response boundary. For item \(i\), choose \(A_i>0\), and let
\begin{equation}\label{eq:rt-exit}
        \tau_i=\inf\{t\ge0:|Z_t^{ip}|=A_i\},
        \qquad
        Y_i=\mathbf 1_{\{Z_{\tau_i}^{ip}=A_i\}}.
\end{equation}
The observed response time is
\begin{equation}\label{eq:rt-transform}
        R_i=t_0+\kappa_i\tau_i,
        \qquad t_0\ge0,
        \quad \kappa_i>0.
\end{equation}
Here \(t_0\) is a non-decision time and \(\kappa_i\) is an item-specific time scale. The observed response time has two components. The decision time $\kappa_i \tau_i$ depends on the person and the item through $\mu_{ip}$ and $A_i$, whereas $t_0$ is common to all responses. The time scale $\kappa_i$ converts the time taken to reach a boundary into units of observed time. Like the time intensity parameter of the lognormal model \citep{van2006lognormal}, it rescales the response time distribution of an item without changing the probability of a correct response. The evidence threshold $A_i$, in contrast, affects both the probability of a correct response and the response time distribution. Because $\sigma$ enters the response time distribution only through $\kappa_i/\sigma^2$ (Figure~\ref{fig:rt-parameters}), $\sigma$ is fixed at one for identification.

\begin{figure}[!htbp]
\centering
\includegraphics[width=\textwidth]{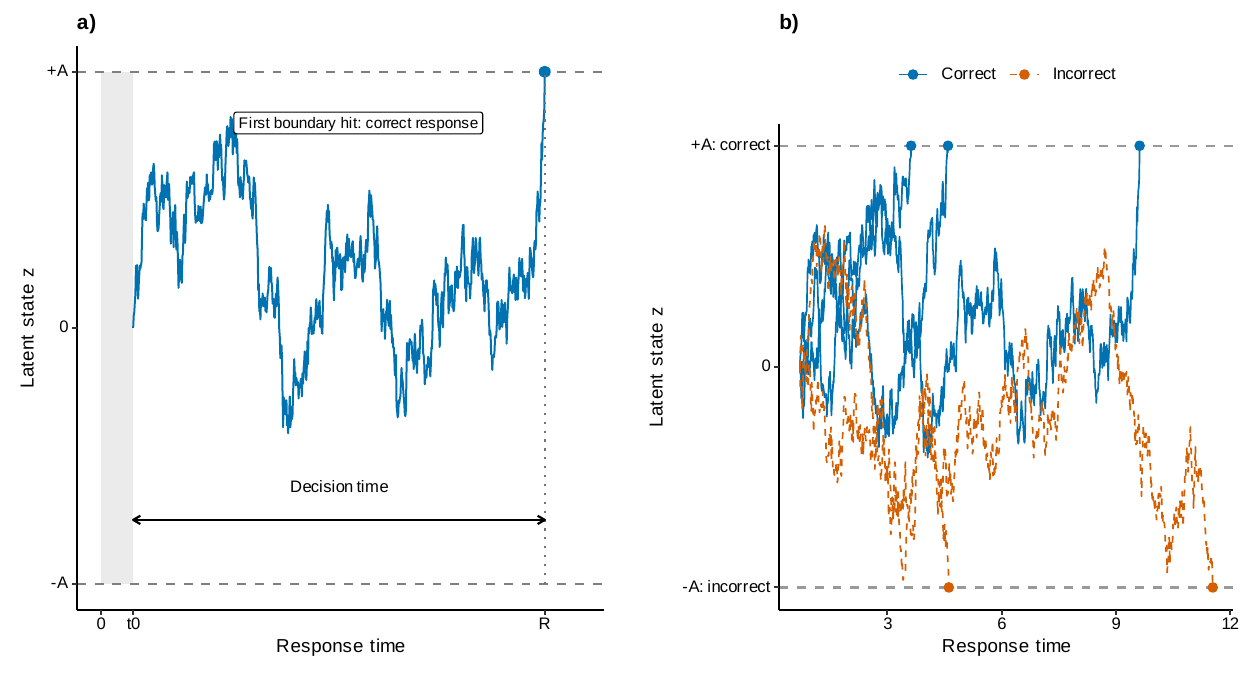}
\caption{(a) Non-decision time (shading) and decision time (arrow) for one correct response. (b) Simulated response times for correct and incorrect answers at the same ability and item difficulty. Both panels use $\mu=1.2$, $A=2$, $\sigma=1$, $\kappa=1$, and $t_0=0.7$.}
\label{fig:response-paths}
\end{figure}

Figure \ref{fig:response-paths}a shows how much of the recorded response time is spent accumulating evidence (the arrow) and how much is non-decision time (the shaded interval). In panel b, even the correct responses to the same item take different amounts of time at the same ability level because the evidence accumulation process follows a different path on each occasion.

\subsubsection{Probability of a correct response}\label{sec:correct-probability}

By Theorem \ref{thm:exit-prob}, for \(x\in(-A_i,A_i)\),
\begin{equation}\label{eq:finite-choice-prob}
        \Prob_x(Y_i=1)
        =\frac{S_{\mu_{ip}}(x)-S_{\mu_{ip}}(-A_i)}
        {S_{\mu_{ip}}(A_i)-S_{\mu_{ip}}(-A_i)},
\end{equation}
where
\[
        S_\mu(x)=x-\mu+\sinh(x-\mu).
\]
Equivalently,
\begin{equation}\label{eq:finite-choice-prob-expanded}
        \Prob_x(Y_i=1)
        =
        \frac{x+A_i+\sinh(x-\mu_{ip})+\sinh(A_i+\mu_{ip})}
        {2A_i+\sinh(A_i-\mu_{ip})+\sinh(A_i+\mu_{ip})}.
\end{equation}

Equation~\eqref{eq:finite-choice-prob-expanded} depends on ability, difficulty and discrimination only through $\mu_{ip} = a_i(\theta_p - \beta_i)$, as the 2PL does. With $x = 0$, the probability of a correct response equals one half when $\theta_p = \beta_i$, and it increases with ability whenever $a_i > 0$. Difficulty therefore retains its interpretation as the location of the item response function, and discrimination continues to govern how sharply the probability changes with ability near $\beta_i$. Differentiating Equation~\eqref{eq:finite-choice-prob-expanded}, we obtain
\begin{equation}
    \left.
    \frac{\partial}{\partial\theta_p}\Prob_0(Y_i=1)
    \right|_{\theta_p=\beta_i}
    =
    \frac{a_i\bigl(\cosh(A_i)-1\bigr)}
         {2\bigl(A_i+\sinh(A_i)\bigr)}.
\end{equation}
For the stationary 2PL model, the slope at $\theta_p=\beta_i$ is $a_i/4$.

Under the two-boundary rule, the slope at the item difficulty depends on the evidence threshold as well as on discrimination. The factor $(\cosh A_i - 1)/(A_i + \sinh A_i)$ increases with $A_i$, so for the same $a_i$ an item with a lower evidence threshold has a flatter response function near its difficulty. Correct responses then provide less evidence that ability exceeds difficulty. If responses follow the two-boundary rule, a 2PL fitted to accuracy alone would estimate a slope that reflects both $a_i$ and $A_i$, and response times are needed to separate the two. This is the mechanism behind the result in \textit{Study 3} that a lower evidence threshold reduced the accuracy of ability estimates more than halving the test length.

\begin{figure}[!htbp]
\centering
\includegraphics[width=\textwidth]{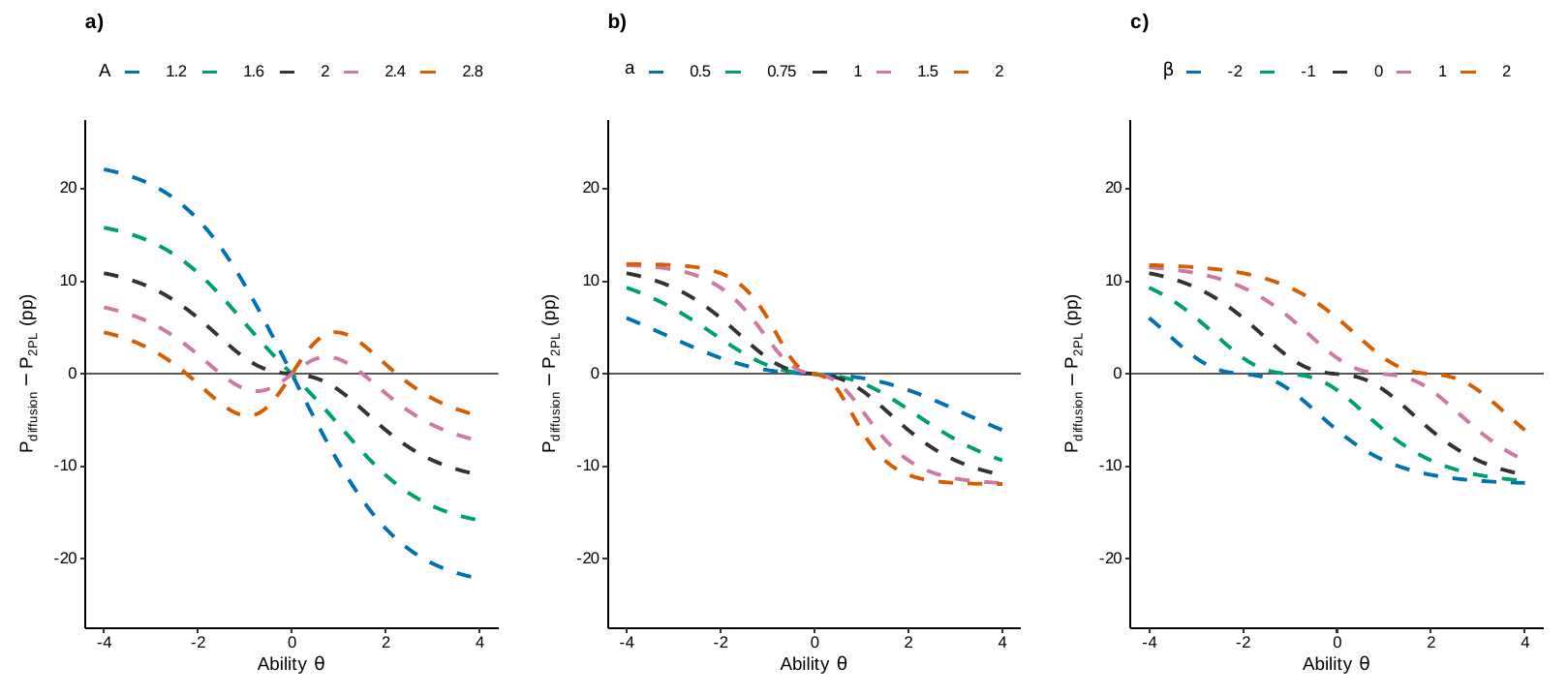}
\caption{Difference in correct-response probability (diffusion minus stationary 2PL), in percentage points, as (a) threshold $A$, (b) discrimination $a$, and (c) difficulty $\beta$ vary. Both models use the same item parameters $a$ and $\beta$. Unless varied, $A=2$, $a=1$, and $\beta=0$.}
\label{fig:accuracy-differences}
\end{figure}

With an evidence threshold of $A=1.2$, the diffusion model predicts a higher probability of a correct response than the stationary 2PL model for respondents whose ability is below item difficulty, and a lower probability for those whose ability is above it (Figure \ref{fig:accuracy-differences}a). At this threshold a correct answer offers less evidence of ability above item difficulty than under the stationary 2PL. In panel b ($A=2$), both models predict a steeper rise in success probability near $\beta_i$ for more discriminating items. For a person one ability unit above $\beta_i$, raising $a_i$ from 0.5 to 2 increases the 2PL probability more than the diffusion probability. Panel c shows the effect of changing difficulty. The difference curve moves along the ability scale, crossing zero at the new difficulty, where both probabilities equal 0.5.

\subsubsection{Response time distributions}\label{sec:response-time-distributions}

The joint distribution can be used to obtain the response time distribution for all responses and the conditional distributions for correct and incorrect responses. The corresponding transforms use
\[
        u_\lambda(x)=\E_x[e^{-\lambda\tau_i}],
        \qquad
        u_\lambda^+(x)=\E_x[e^{-\lambda\tau_i}\mathbf 1_{\{Y_i=1\}}],
        \qquad
        u_\lambda^-(x)=\E_x[e^{-\lambda\tau_i}\mathbf 1_{\{Y_i=0\}}].
\]
Then
\begin{equation}\label{eq:laplace-rt}
        \E_x[e^{-sR_i}]=e^{-st_0}u_{s\kappa_i}(x),
\end{equation}
and, whenever the conditioning event has positive probability,
\begin{equation}\label{eq:conditional-laplace-rt}
        \E_x[e^{-sR_i}\mid Y_i=1]
        =e^{-st_0}\frac{u_{s\kappa_i}^+(x)}{u_0^+(x)},
        \qquad
        \E_x[e^{-sR_i}\mid Y_i=0]
        =e^{-st_0}\frac{u_{s\kappa_i}^-(x)}{u_0^-(x)}.
\end{equation}
The density formula follows directly from Theorem \ref{thm:joint-exit}. If \(f_{\tau_i,Y_i}\) denotes the joint density of \((\tau_i,Y_i)\) obtained from \eqref{eq:exit-density-upper}-\eqref{eq:exit-density-lower} with \(\ell=-A_i\), \(r=A_i\), and \(\mu=\mu_{ip}\), then for \(r>t_0\),
\begin{equation}\label{eq:rt-density}
        f_{R_i,Y_i}(r,y\mid x)
        =\frac{1}{\kappa_i}
        f_{\tau_i,Y_i}\left(\frac{r-t_0}{\kappa_i},y\mid x\right),
        \qquad y\in\{0,1\},
\end{equation}
while \(f_{R_i,Y_i}(r,y\mid x)=0\) for \(r\le t_0\). Therefore
\begin{equation}\label{eq:rt-density-marginal}
        f_{R_i}(r\mid x)
        =\frac{1}{\kappa_i}
        f_{\tau_i}\left(\frac{r-t_0}{\kappa_i}\mid x\right)\mathbf 1_{\{r>t_0\}}.
\end{equation}

\begin{figure}[!htbp]
\centering
\includegraphics[width=\textwidth]{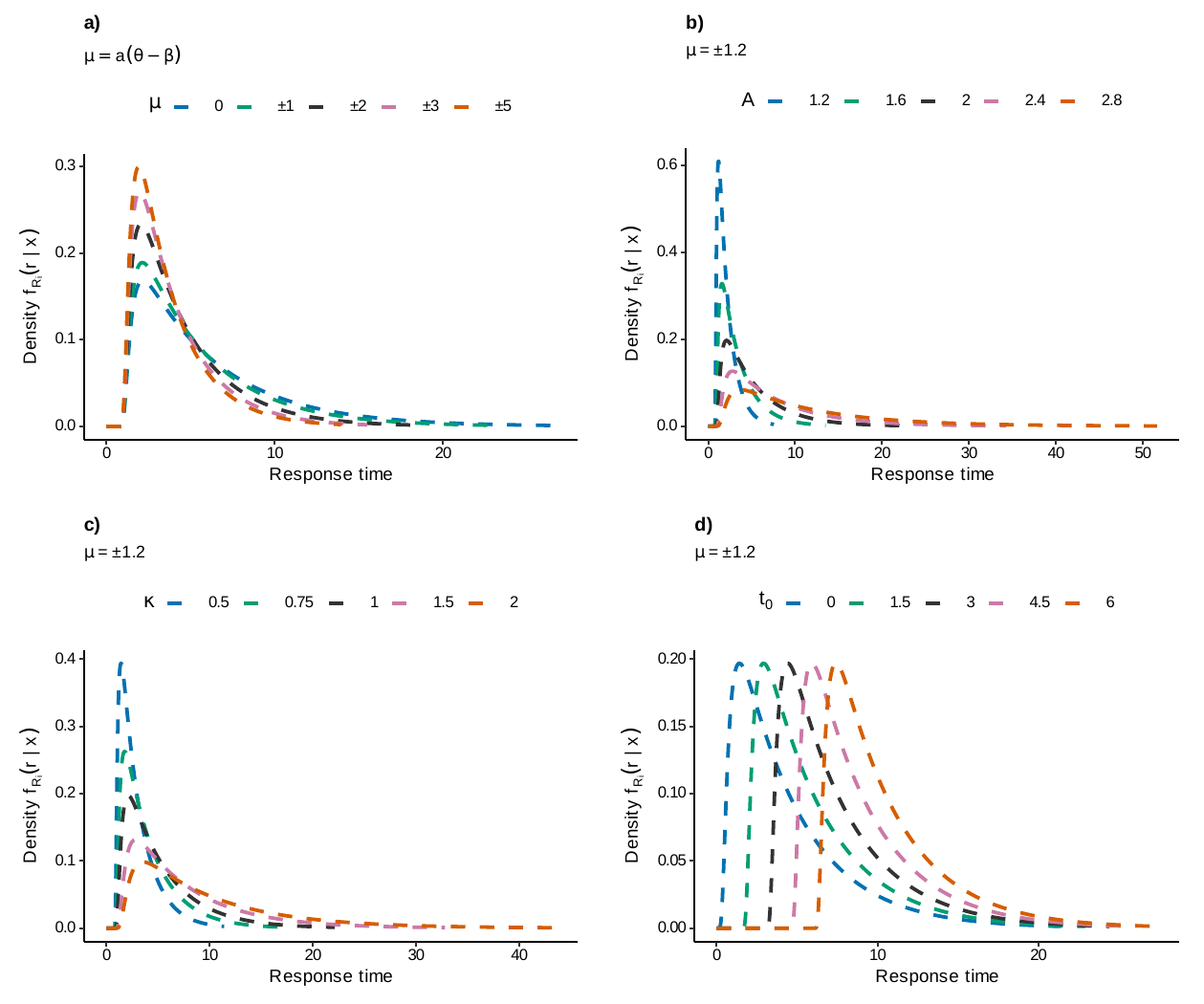}
\caption{Response time densities as (a) $\mu=a(\theta-\beta)$, (b) threshold $A$, (c) time scale $\kappa$, and (d) non-decision time $t_0$ vary. Opposite signs of $\mu$ give the same density, and the effect of $\sigma$ is captured by the time scale $\kappa/\sigma^2$. Unless varied, $|\mu|=1.2$, $A=2$, $\kappa=1$, $t_0=0.7$, and $\sigma=1$.}
\label{fig:rt-parameters}
\end{figure}

The conditional response time distributions allow us to compare the timing of correct and incorrect responses at a given ability level. For fixed person and item parameters and initial state $x$, these densities are
\begin{equation}
    f_{R_i\mid Y_i}(r\mid y,x)
    =
    \frac{f_{R_i,Y_i}(r,y\mid x)}
         {\Prob_x(Y_i=y)},
    \qquad y\in\{0,1\}.
\end{equation}
Each conditional density integrates to one, so their comparison concerns the timing of responses within each outcome category. The above equations show that the time scale and the non-decision time act on the response time distribution as a change of scale and a change of location, and neither affects the probability of a correct response. Response times therefore inform $\kappa_i$ and $t_0$, while the evidence threshold is informed by both responses and response times. Figure~\ref{fig:rt-parameters}a shows a further property that respondents equally far above and below an item difficulty have the same response time distribution, because reversing the sign of $\mu_{ip}$ interchanges the two boundaries. The response time distribution is informative about $|\theta_p - \beta_i|$ but not about its sign, which is informed by the response. A related distance--difficulty relationship has been proposed for personality items \citep{ferrando2007measurement}; in the present model it is a consequence of the response process.

Figure \ref{fig:rt-parameters} pools correct and incorrect answers. In panel a, students equally far above and below the item's difficulty have the same response time distribution because reversing the direction of the evidence interchanges the two response boundaries. In panel b the paths have farther to travel as the boundary widens. In panels c and d the response rule is fixed. Multiplying $\kappa$ by two doubles the time spent reaching the boundary, whereas adding one unit to $t_0$ shifts the entire density one unit to the right.

\subsection{Estimation of item parameters and person ability}\label{sec:estimation}

We first calibrate the items using answers and response times from all respondents. We then hold the item parameters at their estimates and use each respondent's own answers and times to estimate their ability.

We assume a standard normal ability distribution, $\theta_p\sim N(0,1)$, which fixes the origin and unit of the ability scale. For estimation, the initial state is fixed at zero and $\sigma=1$. Discrimination $a_i$, difficulty $\beta_i$, and time scale $\kappa_i$ vary across items. The simulation studies use a common evidence threshold $A_i=A$ and a common non-decision time $t_0$. We estimate the parameters of the basic model, denoted collectively by $\boldsymbol{\psi}$, by maximizing the marginal likelihood
\begin{equation}\label{eq:calibration-likelihood}
    L(\boldsymbol{\psi}) =
    \prod_{p=1}^{N} \int_{\mathbb R}
    \left\{\prod_{i\in\mathcal O_p}
    g_i(r_{pi},y_{pi}\mid\theta,\boldsymbol{\psi})\right\}
    \phi(\theta) d\theta .
\end{equation}
Ability is unobserved, so calibration averages each respondent's likelihood over the standard normal ability distribution using Gauss--Hermite quadrature. Given ability, the basic model treats items as independent and uses the joint density in \eqref{eq:rt-density} for each item with a recorded answer and time ($i\in\mathcal O_p$). In the PISA model (Section \ref{sec:empirical}), students of the same ability can differ in speed across items. Item responses are then conditionally independent given both ability and speed, and calibration averages over their joint distribution. Under the time transformation in \eqref{eq:pisa-rt}, an observed time $r>0$ at speed $s$ corresponds to a first passage time $u=(re^s/\kappa_i)^{1/\gamma_i}$. The joint density for the observed answer and time is the joint density for the answer and first passage time at $u$, multiplied by $u/(\gamma_i r)$.

The expected a posteriori (EAP) score is the mean of the respondent's posterior ability distribution. Two persons with the same answers may consequently receive different scores if their times are more probable at different ability levels. The estimation procedure is the two-stage approach common in operational IRT modeling. Items are calibrated on all respondents, and persons are then scored with the item parameters fixed at their estimates. Response times contribute to ability estimates because they are informative about the distance between ability and item difficulty. Two persons with the same answers may therefore receive different EAP scores. The posterior intervals treat the item parameter estimates as known, so calibration error is not propagated to the ability estimates \citep{cheng2010impact}. This matters for the coverage results of \textit{Study 1}. 

\section{Evaluation of the model}\label{sec:evaluation}

We evaluate what response times add to the measurement of ability and how well the logistic diffusion IRT model accounts for differences in accuracy and response time across items. The three simulation studies address three questions. Study 1 examines whether the item parameters of the logistic diffusion IRT model can be recovered and whether adding response times to answers improves ability estimates. Study 2 examines how ability estimates depend on the model used to calibrate the items when the data are generated by a different model. Study 3 examines how ability estimates change when response behaviour departs from the model or when fewer items are observed. In three simulation studies, the item parameters and abilities are known, so we can assess how accurately they are estimated and how ability estimates change when the assumptions about responding no longer hold. We then analyze student responses to a PISA mathematics booklet. We examine whether items of similar difficulty take different amounts of time, whether the relationship between response time and accuracy varies with student ability, and how well the model predicts each student's answer and response time on a withheld item.

\subsection{Simulation studies}\label{sec:simulation}

In Study 1, we assess how accurately the logistic diffusion IRT model recovers the item parameters. We then score each respondent twice using the same item estimates to determine whether adding response times to answers improves ability estimates and how it affects interval coverage. In Studies 2 and 3, answers and times both contribute to item calibration, but ability scores use answers alone. Study 2 compares five models when each in turn generates the data. Study 3 examines how ability recovery changes when respondents guess, take longer, or answer with less evidence, and when fewer items are observed. The simulation studies took about 15 hours to run on a 24 core machine with parallelized R code.

\paragraph{Study 1: Recovery of item parameters and ability.} Study 1 crossed two sample sizes (500 and 1,000 persons), two test lengths (20 and 30 items), two evidence thresholds (A = 2.2 and 3.7) and three compositions of item difficulty, leading to 24 design conditions. We generated 20 data sets in each of the 24 settings and fitted the logistic diffusion IRT model to all 480.

The sample sizes range in which the typical 2PL is usually calibrated. The logistic diffusion IRT model has more item parameters than the 2PL, but each response also contributes a response time, so 500 persons is a demanding but realistic lower level. It also matches the number of students in the PISA example. Doubling the sample to 1,000 shows whether estimation errors are due to sampling variability, in which case they should shrink, or to systematic sources, in which case they should not. The test lengths of 20 and 30 items reflect the instruments for which joint models are typically used. Test length matters for scoring rather than calibration, because it determines how much information each person contributes to their own ability estimate, and therefore how much response times can add. The two evidence thresholds were chosen to represent two qualitatively different response functions. At A = 2.2, the item response function is close to the stationary 2PL while at A = 3.7 the response function is markedly steeper, and response times are longer and more dispersed (simiilar to the PISA example). 

The three compositions of item difficulty---easy, moderate and difficult items---had difficulties drawn uniformly from $[-1.5, -0.5]$, $[-0.5, 0.5]$ and $[0.5, 1.5]$. The balanced form, with approximately equal numbers in each band, spreads information across the ability distribution. The form with about half difficult items represents a test targeting above the average respondent, as in selection testing, and the form with about half moderate items concentrates information near the mean. A mostly easy form was not included because, with symmetric boundaries and a standard normal ability distribution, it is equivalent to the mostly difficult form after reversing the sign of ability and difficulty and relabelling the responses.

Item discriminations were drawn uniformly from [0.5, 4], so that recovery could be examined separately for moderately, highly and very highly discriminating items (Table~\ref{tab:joint-recovery}). The upper part of this range exceeds the values usually estimated for operational items. It was included deliberately, to locate where recovery deteriorates. Time scales were drawn from [1.5, 3.5], so items differed in time demand by up to a factor of about 2.3 independently of their difficulty. The non-decision time was $t_0 = 0.2$, small relative to typical decision times, with $\sigma =1$ and an initial state of zero.

\subparagraph{Outcomes} Item parameter recovery was assessed by bias and RMSE. For ability, each person was scored twice with the same item parameter estimates: (1) from answers alone and (2) from answers and response times. Holding the item parameter estimates fixed, two EAP scores for each person was calculated, one from their answers alone and one from their answers and response times. For each scoring method, the proportion of respondents whose nominal 95\% posterior interval contained their true ability was recorded. Differences between the two scoring methods were computed within each replication and averaged over the 24 equally weighted conditions. Monte Carlo standard errors are provided in parentheses. For the pooled bias and RMSE in Tables \ref{tab:joint-recovery} and \ref{tab:discrimination-recovery}, these were calculated by the delta method across independent data sets within each setting, accounting for dependence among item estimates from the same data set.

\subparagraph{Results} Item difficulty, time scale, evidence threshold and non-decision time were recovered with small bias (Table \ref{tab:joint-recovery}). Discrimination was underestimated by 0.305 on average, and the underestimation increased with the generating discrimination (Table \ref{tab:discrimination-recovery}). For items with $3 \leq a_i \leq 4$, doubling the sample size reduced RMSE from 0.850 to 0.730, but the estimates remained 0.573 below the generating values on average in the larger sample. Because the bias did not decrease with sample size, it is unlikely to be a small-sample effect. A plausible source is the numerical approximation used in calibration, since steep response functions are where quadrature over ability and truncation of the eigenfunction expansion are least accurate.

\begin{table}[!htbp]
\centering
\caption{Parameter recovery for the logistic diffusion IRT model in Study 1. Monte Carlo standard errors are in parentheses.}
\label{tab:joint-recovery}
\begin{tabular}{lrr}
\toprule
Parameter & Bias & RMSE\\
\midrule
Item discrimination $a_i$ & $-0.305$ (0.0055) & $0.531$ (0.0076)\\
Item difficulty $\beta_i$ & $0.003$ (0.0050) & $0.193$ (0.0030)\\
Item time scale $\kappa_i$ & $-0.041$ (0.0035) & $0.140$ (0.0025)\\
Response boundary $A$ & $0.010$ (0.0014) & $0.032$ (0.0014)\\
Non-decision time $t_0$ & $0.024$ (0.0026) & $0.063$ (0.0025)\\
\bottomrule
\end{tabular}
\end{table}

\begin{figure}[!htb]
\centering
\includegraphics[width=\linewidth]{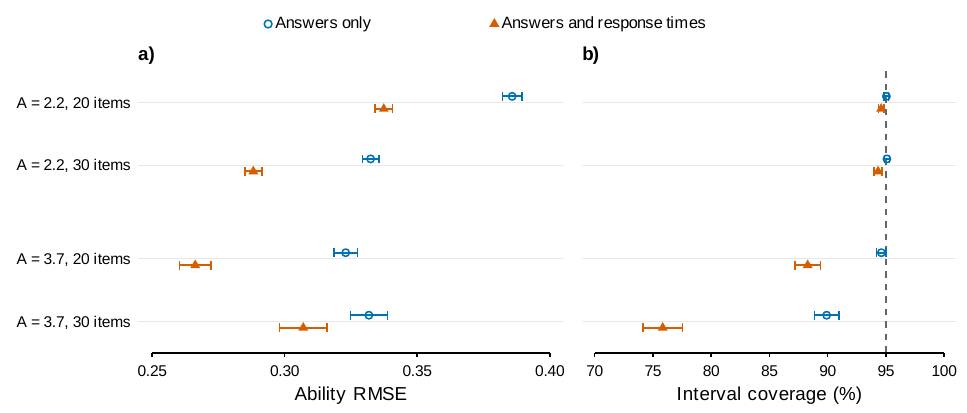}
\caption{Ability RMSE for Study 1 (a) and coverage of nominal 95\% ability intervals (b) for the same respondents scored with answers alone or with answers and response times. Points are means over 120 data sets for each combination of boundary and test length. The whiskers give approximate 95\% Monte Carlo intervals for these means. The dashed line indicates nominal coverage.}
\label{fig:simulation-precision}
\end{figure}

Adding response times to answers reduced mean ability RMSE from $0.343$ to $0.300$, a reduction of $0.0435$ ($0.0006$), or $12.5\%$. Coverage of the nominal $95\%$ intervals fell from $93.7\%$ to $88.3\%$, a reduction of $5.40$ ($0.15$) percentage points. The loss of coverage depended on the evidence threshold and test length (Figure~\ref{fig:simulation-precision}). At $A = 2.2$, joint scoring gave coverage of $94.6\%$ with $20$ items and $94.3\%$ with $30$ items. With $30$ items and $A = 3.7$, coverage fell from $89.9\%$ with answers alone to $75.8\%$ with answers and times.

The pattern is consistent with two features of the model and the estimation procedure. At the higher threshold, response time distributions are more dispersed and depend more strongly on $|\theta_p - \beta_i|$ (Figure~\ref{fig:rt-parameters}), so response times sharpen the posterior distribution of ability. The posterior intervals, however, treat the item parameter estimates as known. As the information contributed by each person increases, calibration error becomes a larger share of the total error, and intervals that ignore it become too narrow \citep{cheng2010impact}. The underestimation of high discriminations may contribute as well.

\begin{table}[!htbp]
\centering
\caption{Recovery of item discrimination in Study 1 by generating discrimination and number of respondents. Monte Carlo standard errors are in parentheses.}
\label{tab:discrimination-recovery}
\small
\setlength{\tabcolsep}{4pt}
\begin{tabular}{lrrrr}
\toprule
 & \multicolumn{2}{c}{500 respondents} & \multicolumn{2}{c}{1,000 respondents} \\
\cmidrule(lr){2-3}\cmidrule(lr){4-5}
Generating discrimination & Bias & RMSE & Bias & RMSE \\
\midrule
$0.5 \leq a_i < 1$ & $-0.076$ (0.0047) & $0.137$ (0.0031) & $-0.084$ (0.0033) & $0.121$ (0.0027) \\
$1 \leq a_i < 2$ & $-0.159$ (0.0067) & $0.291$ (0.0125) & $-0.181$ (0.0055) & $0.257$ (0.0042) \\
$2 \leq a_i < 3$ & $-0.292$ (0.0136) & $0.550$ (0.0244) & $-0.343$ (0.0085) & $0.471$ (0.0070) \\
$3 \leq a_i \leq 4$ & $-0.481$ (0.0185) & $0.850$ (0.0213) & $-0.573$ (0.0124) & $0.730$ (0.0091) \\
\bottomrule
\end{tabular}
\end{table}

\paragraph{Study 2: Ability recovery under different response models.} In Study 2, we consider the models that represent the main families of joint models for accuracy and response time distinguished by \citet{de2019overview} along with the logistic diffusion model. The hierarchical model (HM) combines a 2PL response function with lognormal response times \citep{van2007hierarchical} and is the most widely used joint model \citep{bolsinova2017conditional}. The response accuracy and response time models are conditionally independent given ability $\theta_p$ and speed $\tau_p$.

\[P(Y_{pi}=1\mid\theta_p)=\frac{1}{1+\exp\{-a_i(\theta_p-\beta_i)\}},\qquad \log R_{pi}\mid\tau_p\sim N\!\left(\lambda_i-\tau_p,\ \phi_i^{-2}\right),\]

\[f(r\mid\tau_p)=\frac{\phi_i}{r\sqrt{2\pi}}\exp\!\left[-\frac{\phi_i^2}{2}\{\log r-(\lambda_i-\tau_p)\}^2\right],\qquad (\theta_p,\tau_p)^\top\sim N_2\!\left(\mathbf{0},\begin{pmatrix}1&\rho\,\sigma_\tau\\ \rho\,\sigma_\tau&\sigma_\tau^2\end{pmatrix}\right).\]

Here $\lambda_i$ is the item time intensity and $\phi_i$ the time discrimination. The drift diffusion model (DDM) follows \citet{ratcliff2016diffusion}. The evidence follows a Wiener process with constant drift between absorbing boundaries at 0 and $\alpha_i$, starting midway:

\[dX_t=v_{pi}\,dt+dW_t,\qquad v_{pi}=\theta_p-\beta_i,\qquad X_0=\alpha_i/2,\qquad Y_{pi}=\mathbf{1}\{X_{D}=\alpha_i\},\qquad R_{pi}=t_0+D_{pi},\]

where $D_{pi}$ is the first exit time from (0, $\alpha_i$). The response probability is a 2PL with discrimination equal to the boundary separation,

\[P(Y_{pi}=1\mid\theta_p)=\frac{1}{1+\exp\{-\alpha_i(\theta_p-\beta_i)\}},\]

and the decision time density at the lower boundary is the Wiener first passage time density \citep{ratcliff1978theory, navarro2009fast}, with w = 1/2 the relative starting point:
\[f_{D,Y}(t,0)=\frac{\pi}{\alpha_i^2}\exp\!\left(-v_{pi}\alpha_i w-\frac{v_{pi}^2t}{2}\right)\sum_{m=1}^{\infty}m\,\sin(m\pi w)\exp\!\left(-\frac{m^2\pi^2t}{2\alpha_i^2}\right).\]

The density at the upper boundary is obtained by replacing $v_{pi}$ with $-v_{pi}$ and $w$ with $1-w$.

The Birnbaum--Saunders (BS) race model (\citet{ranger2015race}) has two accumulators, one for the correct and one for the incorrect response, have finishing times $T_1$ and $T_0$, and the first to finish determines the response.
\[Y_{pi}=\mathbf{1}\{T_{1,pi}<T_{0,pi}\},\qquad R_{pi}=t_0+\min(T_{1,pi},T_{0,pi}).\]

Each finishing time has a Birnbaum–Saunders distribution with shape $\nu$ scale $\eta_{k,pi}$

\[F_k(t)=\Phi\!\left[\frac{1}{\nu}\left(\sqrt{\frac{t}{\eta_{k,pi}}}-\sqrt{\frac{\eta_{k,pi}}{t}}\right)\right],\] \[f_k(t)=\frac{1}{2\nu t}\left(\sqrt{\frac{t}{\eta_{k,pi}}}+\sqrt{\frac{\eta_{k,pi}}{t}}\right)\phi\!\left[\frac{1}{\nu}\left(\sqrt{\frac{t}{\eta_{k,pi}}}-\sqrt{\frac{\eta_{k,pi}}{t}}\right)\right].\]
The joint density of response accuracy and time is given by-
\[f_{R,Y}(r,1)=\int_{r-t_0}^{\infty} f_{T_1,T_0}(r-t_0,s)\,ds,\qquad f_{R,Y}(r,0)=\int_{r-t_0}^{\infty} f_{T_1,T_0}(s,r-t_0)\,ds,\]

which reduces to $f_1(r - t_0)\{1 - F_0(r - t_0)\}$ and $f_0(r - t_0)\{1 - F_1(r - t_0)\}$ when the accumulators are independent.

The Weibull race (WR) model \citep{tuerlinckx2005two} involve two independent decision nodes each having Weibull finishing times with a common shape $k_i$.
\[P(T_{1,pi}>t)=\exp\!\left\{-\lambda_i N^{a_i/2}e^{\,a_i(\theta_p-\beta_i)/2}\,t^{k_i}\right\},\] \[P(T_{0,pi}>t)=\exp\!\left\{-\lambda_i N^{a_i/2}e^{-a_i(\theta_p-\beta_i)/2}\,t^{k_i}\right\},\]

with N fixed (N = 10) for identification. The response probability is the 2PL, and the response and the response time are independent given the parameters.

For the simulation, each of five models generated 20 data sets of 1,000 respondents answering a test of 20 items. Item discriminations ranged from 0.7 to 1.5 and $A=2.5$, $t_0=0.2$ for the logistic diffusion IRT model. We fitted all five models to each data set. 

\subparagraph{Outcomes} The outcome was the accuracy of ability estimates. In each data set, every fitted model was calibrated on the answers and response times of all respondents. Each respondent was then scored by the EAP estimate of ability given their answers alone, with the item parameters fixed at that model estimates. All five models were applied to the same respondents and the same answers, so differences between their ability estimates arise only from the item response functions estimated at calibration. For each combination of generating and fitted model, we computed the RMSE of the EAP estimates against the generating abilities within each data set and averaged it over the 20 data sets. In Table \ref{tab:simulation-ability}, the Monte Carlo standard error of each mean is the standard deviation of the 20 replication-level RMSEs divided by $\sqrt{20}$.

\subparagraph{Results}Table \ref{tab:simulation-ability} reports the mean ability RMSE within each setting. The logistic diffusion IRT model had lower RMSE than the hierarchical model when the logistic diffusion (0.448 versus 0.461) or the Birnbaum–Saunders race (0.347 versus 0.449) generated the data. When the hierarchical model generated the data, person speed varied independently of ability, and the hierarchical model gave more accurate scores (0.452 versus 0.487). Under data generated by the DDM and the Weibull race, the differences between models were small. The choice of calibration model can therefore affect ability estimates even when response times are not used in scoring. Across the five generating models, the largest increase in RMSE of the logistic diffusion IRT model over the best-performing model in the same row was 0.095, compared with 0.197 for the hierarchical model and 0.018 for the Birnbaum–Saunders race model. The choice of model for calibration can thus affect ability estimates even when times are used only during calibration.

\begin{table}[!htbp]
\centering
\caption{Ability RMSE in Studies 2 and 3. Columns identify the model used to estimate ability for the same respondents in each row. Monte Carlo standard errors from the 20 replications are in parentheses below each mean. The model proposed in this paper appears in bold.}
\label{tab:simulation-ability}
\small
\setlength{\tabcolsep}{5pt}
\renewcommand{\arraystretch}{1.3}
\begin{tabular}{lrrrrr}
\toprule
 & \shortstack{\textbf{Logistic}\\\textbf{diffusion}} & HM & DDM & BS & WR\\
\midrule
\multicolumn{6}{l}{\textit{Study 2. Generating model}}\\
\textbf{Logistic diffusion} & \shortstack[r]{0.448\\(0.0031)} & \shortstack[r]{0.461\\(0.0032)} & \shortstack[r]{0.476\\(0.0033)} & \shortstack[r]{0.466\\(0.0033)} & \shortstack[r]{0.465\\(0.0033)}\\[3pt]
Hierarchical model (HM) & \shortstack[r]{0.487\\(0.0039)} & \shortstack[r]{0.452\\(0.0030)} & \shortstack[r]{0.450\\(0.0038)} & \shortstack[r]{0.441\\(0.0032)} & \shortstack[r]{0.472\\(0.0044)}\\[3pt]
Drift diffusion model (DDM) & \shortstack[r]{0.358\\(0.0025)} & \shortstack[r]{0.355\\(0.0022)} & \shortstack[r]{0.365\\(0.0032)} & \shortstack[r]{0.346\\(0.0022)} & \shortstack[r]{0.376\\(0.0029)}\\[3pt]
Birnbaum--Saunders race model (BS) & \shortstack[r]{0.347\\(0.0054)} & \shortstack[r]{0.449\\(0.0084)} & \shortstack[r]{0.522\\(0.0091)} & \shortstack[r]{0.252\\(0.0028)} & \shortstack[r]{0.587\\(0.0087)}\\[3pt]
Weibull race model (WR) & \shortstack[r]{0.452\\(0.0032)} & \shortstack[r]{0.434\\(0.0029)} & \shortstack[r]{0.433\\(0.0029)} & \shortstack[r]{0.434\\(0.0028)} & \shortstack[r]{0.434\\(0.0029)}\\[3pt]
\midrule
\multicolumn{6}{l}{\textit{Study 3. Logistic diffusion generating settings}}\\
Unmodified responses & \shortstack[r]{0.375\\(0.0027)} & \shortstack[r]{0.385\\(0.0025)} & \shortstack[r]{0.433\\(0.0030)} & \shortstack[r]{0.391\\(0.0026)} & \shortstack[r]{0.395\\(0.0025)}\\[3pt]
Long times (1\%) & \shortstack[r]{0.372\\(0.0026)} & \shortstack[r]{0.382\\(0.0025)} & \shortstack[r]{0.435\\(0.0024)} & \shortstack[r]{0.387\\(0.0025)} & \shortstack[r]{0.392\\(0.0018)}\\[3pt]
Long times (5\%) & \shortstack[r]{0.375\\(0.0030)} & \shortstack[r]{0.386\\(0.0030)} & \shortstack[r]{0.432\\(0.0038)} & \shortstack[r]{0.392\\(0.0033)} & \shortstack[r]{0.397\\(0.0026)}\\[3pt]
Rapid guessing (5\%) & \shortstack[r]{0.403\\(0.0031)} & \shortstack[r]{0.412\\(0.0031)} & \shortstack[r]{0.442\\(0.0029)} & \shortstack[r]{0.420\\(0.0031)} & \shortstack[r]{0.417\\(0.0029)}\\[3pt]
10 of 20 items observed & \shortstack[r]{0.498\\(0.0041)} & \shortstack[r]{0.506\\(0.0041)} & \shortstack[r]{0.530\\(0.0041)} & \shortstack[r]{0.511\\(0.0041)} & \shortstack[r]{0.509\\(0.0039)}\\[3pt]
Shorter decision times & \shortstack[r]{0.376\\(0.0029)} & \shortstack[r]{0.386\\(0.0027)} & \shortstack[r]{0.437\\(0.0029)} & \shortstack[r]{0.391\\(0.0028)} & \shortstack[r]{0.399\\(0.0029)}\\[3pt]
Lower response threshold ($A=1.8$) & \shortstack[r]{0.550\\(0.0035)} & \shortstack[r]{0.563\\(0.0032)} & \shortstack[r]{0.562\\(0.0033)} & \shortstack[r]{0.566\\(0.0032)} & \shortstack[r]{0.567\\(0.0036)}\\[3pt]
\bottomrule
\end{tabular}
\end{table}

\paragraph{Study 3: Sensitivity to changes in response behaviour and test length.} In Study 3, we used the logistic diffusion model as the datagenerating model under seven test conditions. We fit all five model described in Study 2 to the generated data. 20 data sets of 1,000 persons answering 20 items were generated for each condition and these conditions were not crossed. In the baseline condition (unmodified responses), the evidence threshold was $A=3.5$, the non-decision time was $t_0 = 0.2$, and all answers and response times were used for calibration and scoring. For each of the six changes (listed below), we recalibrated all five models, scored respondents from their available answers, and compared the scores with their generating abilities. The outcome was the RMSE of the ability estimates against the generating abilities, computed within each data set and averaged over the 20 data sets for each combination of condition and fitted model (Table~\ref{tab:simulation-ability}). Each model was compared with its own baseline RMSE.

\emph{Occasional long times and rapid guesses.} We lengthened a randomly selected 1\% or 5\% of response times by multiplying them by lognormal factors with median 4 and log standard deviation 0.3. Only the times used in calibration changed. At both rates the five models scored respondents about as accurately as at baseline, as measured by mean RMSE. To simulate rapid guessing, we replaced a random 5\% of answers with guesses, each correct with probability 0.5 at any ability. We set the corresponding times to one fifth of their original values, with a minimum of 0.05. Ability RMSE then increased for all five models relative to baseline.

\emph{Fewer observed items.} For each respondent we drew 10 of the 20 items at random and used only the answers and times on those items. With half as many answers per person, mean ability RMSE rose for all five models. For logistic diffusion IRT the increase was from 0.375 to 0.498.

\emph{Faster responses.} We generated shorter response times in two ways. First, we reduced each respondent's evidence accumulation time by 40\% and kept the answer they had given. The resulting time was $t_0+0.60(R-t_0)$, and all five models gave ability RMSEs close to baseline. In the second setting we generated answers and times at $A=1.8$, so respondents needed less evidence to answer. Consider an item with $a_i=1$. A respondent one ability unit above its difficulty had a probability of success of 0.82 at the original threshold and 0.70 at the lower one. One unit below its difficulty, the corresponding probabilities were 0.18 and 0.30. With the lower threshold, a correct answer gave less support for the conclusion that ability exceeded item difficulty. Across all five models, 10 items at $A=3.5$ gave more accurate ability estimates than 20 items at $A=1.8$.

\subsection{Real data analysis of PISA mathematics responses}\label{sec:empirical}

In the 2018 Programme for International Student Assessment (PISA), 500 Canadian students answered the 12 items in mathematics booklet M01. The \texttt{pisaRT} package provides a score and a response time in seconds for each of their 6,000 responses \citep{becker2023pisart}. We scored full credit as correct and all other responses as incorrect. Across items, the proportion correct ranged from 0.036 to 0.872 and the median response time from 38 to 185 seconds. We compared response times across items with similar full-credit rates. We also compared full credit on early and later answers for students above and below the item's difficulty. We assessed prediction by withholding an item and predicting both the student's answer and the time taken.

Some students work faster than others across the booklet. The variation in recorded times also depends on the item. If student $p$ first reaches a response boundary on item $i$ at time $\tau_{pi}$, the model predicts a recorded response time of
\begin{equation}\label{eq:pisa-rt}
 R_{pi}=\kappa_i e^{-s_p}\tau_{pi}^{\gamma_i}, \qquad \gamma_i>0.
\end{equation}
Here $s_p$ is the student's speed across items and $\gamma_i$ governs how much variation in the time taken to reach a boundary appears in the recorded response times for item $i$. Student speed $s_p$ rescales all of a student's response times by $e^{-s_p}$, so students of the same ability can differ in how quickly they work across the booklet. The exponent $\gamma_i$ determines how strongly variation in first passage times is transmitted to recorded response times on item $i$. With $s_p = 0$, $\gamma_i = 1$ and $t_0 = 0$, \eqref{eq:pisa-rt} reduces to the time scaling in~\eqref{eq:rt-transform}.

For the PISA booklet, we estimated $a_i$, $\beta_i$, $A_i$, $\kappa_i$ and $\gamma_i$ for each item and fixed $t_0=0$. We assumed a joint normal distribution for ability and speed, with both means fixed at zero and ability variance fixed at one. We estimated the correlation between ability and speed and the standard deviation of speed. The latter estimate was $0.29$, so for the same item and first passage time, a student one standard deviation above average speed takes about $25\%$ less time than a student at average speed. The estimated correlation between ability and speed was $-0.41$: students of higher ability tended to work more slowly. In a low-stakes assessment, this is consistent with evidence that more skilled students adapt the time they spend to the demands of the task \citep{naumann2019skilled}. In the model, ability is related to response time in two ways: within an item, through the distance between ability and difficulty, and across items, through the correlation between ability and speed.

\begin{figure}[!htb]
\centering
\includegraphics[width=\linewidth]{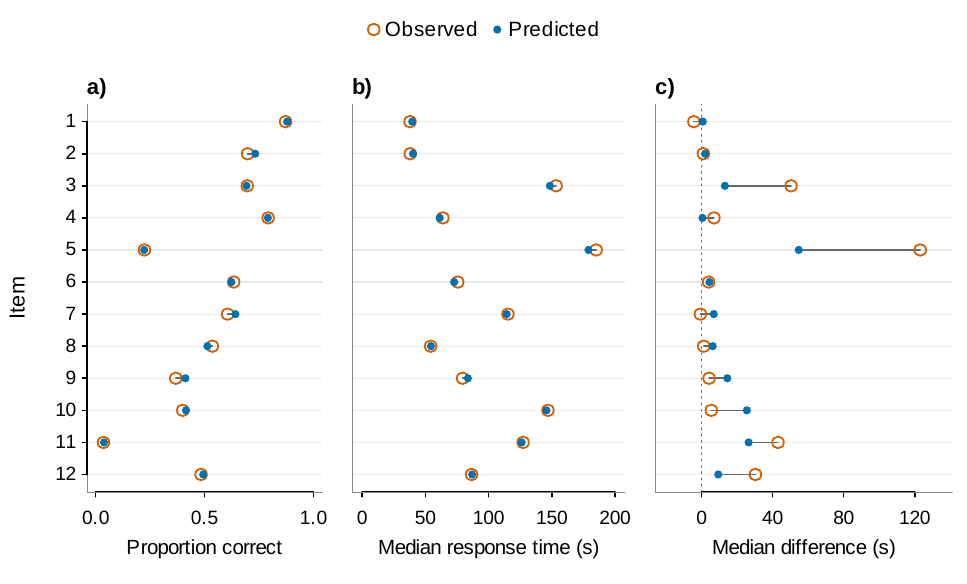}
\caption{Observed and predicted proportion correct (a), median response time (b), and difference between median times for correct and incorrect answers (c) for the 12 PISA items. Predictions average over the joint distribution of student ability and speed. Positive differences indicate longer median times for correct answers.}
\label{fig:pisa-item-fit}
\end{figure}

\paragraph{Item difficulty and time demand.}
About 70\% of students earned full credit on each of items 2 and 3, yet the observed median time was 38 seconds on item 2 and 153 seconds on item 3 (Figure \ref{fig:pisa-item-fit}a,b). We matched the chance of full credit across the two items by evaluating item 2 at its estimated difficulty ($-1.29$) and item 3 at its estimated difficulty ($-0.77$). At those respective ability levels, each student has a 50\% chance of full credit. At average speed, the predicted median times are 45 and 183 seconds. Item 3 thus takes about four times as long even when the items are equally challenging for the students being compared. Its estimated evidence threshold was lower than that of item 2 (1.73 versus 2.20), but its time scale was larger (100 versus 21 seconds). Students need less evidence to respond to item 3, but that evidence accumulates more slowly in observed time. A model for accuracy alone would treat the two items as nearly equivalent; under a time limit, they are not.

\paragraph{Prediction of a rare outcome} Only 18 of the 500 students (3.6\%) earned full credit on item 11. Using each student's answers and response times on other items, the model predicted a 3.7\% full-credit rate for item 11, close to the observed rate.

\paragraph{Short response times and ability}
On item 8 (Table \ref{tab:pisa-item8-conditional}), the model predicts a 20\% chance of full credit for a student one ability unit below its difficulty and an 80\% chance for a student one unit above it. At the same speed, the two students have the same predicted response time distribution. At average speed, the median is 55 seconds for both. A short time alone therefore gives no indication of which of these two students answered the item. The direction is informed by the answer. Averaged over the distribution of speed at each ability, predicted medians increase with ability (Table~\ref{tab:pisa-item8-conditional}) because ability and speed are negatively correlated.

\begin{table}[!htbp]
\centering
\caption{Predictions for PISA item 8 at three points on the standard normal ability scale. The middle row is the difficulty estimated by the logistic diffusion IRT model. Response times are medians in seconds, averaging over the distribution of speed at each ability. The estimated correlation between ability and speed was $-0.41$, so these medians differ from the predictions at fixed speed in the text. The hierarchical model (HM) gives the same median for correct and incorrect responses at a given ability.}
\label{tab:pisa-item8-conditional}
\begin{tabular}{lrrrrr}
\toprule
 & \multicolumn{4}{c}{Logistic diffusion} & HM \\
\cmidrule(lr){2-5}\cmidrule(l){6-6}
Ability & $P(\mathrm{correct})$ & All responses & Correct & Incorrect & All responses \\
\midrule
$-1.05$ & $0.198$ & 47.6 & 47.6 & 47.6 & 44.0 \\
$-0.05$ & $0.500$ & 58.8 & 58.8 & 58.8 & 51.5 \\
$0.95$ & $0.802$ & 60.3 & 60.3 & 60.3 & 60.3 \\
\bottomrule
\end{tabular}
\end{table}

\paragraph{Do early responses earn full credit?}

Section \ref{sec:two-sided} predicts that early responses are less likely than later ones to earn full credit above item difficulty and more likely to do so below it. We tested this prediction on withheld responses to the same item among students of similar ability, estimated from their other answers. We classified a response as early if its time was at or below the item's median in the training data for that fold. Above item difficulty, early responses earned full credit 3.6 percentage points less often than later ones, as predicted. Later responses also earned full credit more often among students whose estimated ability was below item difficulty. The difference was 11.2 percentage points, or 6.8 points when we excluded the fastest 20\% of responses on each item. Using each student's predicted median time on the item to classify a response as early gave the same direction of difference.

Item 5 contributed most to the difference in full-credit rates for this group. It asked students to supply their own answer about population pyramids \citep[Table A.4]{oecd2017pisa2015technical}. Partial credit counted as incorrect in the binary scores \citep{becker2023pisart}. Of the 500 students, 476 had estimated ability below the item's difficulty. After trimming, 10.4\% of their early responses earned full credit, compared with 35.2\% of their later responses. Median response times on this item were 133 seconds in the early group and 295 seconds in the later group. Repeating the trimmed comparison without item 5 reduced the gap from 6.8 to 3.3 percentage points. Whichever item we omitted, later responses earned full credit more often.

Three features of the data that lie outside the model may explain this result. On a constructed-response item, the time spent includes constructing and writing an answer, and with partial credit scored as incorrect, early responses are more likely to be incomplete. Some very fast responses from students well below an item difficulty may reflect disengagement rather than evidence accumulation \cite{wise2005response}; excluding the fastest 20\% of responses reduced the difference by about 40\%. Finally, students may differ in the rate at which they make progress on an item, which in diffusion models produces slow incorrect responses \citep{ratcliff2008diffusion}.

\paragraph{Prediction of withheld responses.}
For each student, we randomly allocated the 12 items to five groups. We withheld two or three items at a time, calibrated each model on the remaining responses, and used the student's answers and times on the other items to predict the withheld answers and times. Each of the 6,000 responses was withheld once. For a student with observed answers and times $O$ and withheld answers and times $H$, the predictive log score is $\log p(H\mid O)$, the log of the joint predictive density evaluated at the responses that student actually gave. We calculated a score per withheld response in each fold by summing the students' log scores and dividing by the number of responses withheld. Table \ref{tab:pisa-prediction} averages these values over the five folds, with higher scores indicating better prediction. Cross-validated log predictive densities have also been used in psychometrics by \citet{garrard2016novel} and \citet{kaplan2021uncertainty}.

HM had the highest predictive log score in every fold and the lowest values of all four information criteria (Table \ref{tab:pisa-prediction}). Logistic diffusion ranked second on prediction. Of HM's advantage of 0.088 log units per withheld response, 0.085 came from predicting time given the answer. One group of three withheld responses, including a time of 0.883 seconds, accounted for about 56\% of the difference between the two mean fold scores. Figure \ref{fig:pisa-item-fit} illustrates a limitation of the logistic diffusion time predictions. The predicted median was within 6.2 seconds of the observed median on every item (panel b), but the model underestimated how much longer full-credit responses took on items 3 and 5 (panel c). On item 5, for example, the observed difference between the two medians was 123 seconds and the model predicted 55 seconds.

\begin{table}[!htbp]
\centering
\caption{Fit and prediction for the Canadian PISA mathematics booklet. The proposed logistic diffusion model (bold) includes person speed and an evidence threshold and time exponent for each item.}\label{tab:pisa-prediction}
\small
\setlength{\tabcolsep}{3pt}
\begin{tabular}{@{}lrrrrrr@{}}
\toprule
Model & Log likelihood & AIC & BIC & CAIC & SABIC & \shortstack{Predictive\\log score} \\
\midrule
\textbf{Logistic diffusion} & $-33791.9$ & $67707.8$ & $67969.1$ & $68031.1$ & $67772.3$ & $-5.670$ \\
HM & $-33786.8$ & $67673.7$ & $67884.4$ & $67934.4$ & $67725.7$ & $-5.582$ \\
DDM & $-68997.2$ & $138066.3$ & $138218.0$ & $138254.0$ & $138103.8$ & $-11.477$ \\
BS & $-34418.4$ & $68934.8$ & $69141.3$ & $69190.3$ & $68985.7$ & $-5.756$ \\
WR & $-35001.7$ & $70059.3$ & $70177.3$ & $70205.3$ & $70088.4$ & $-5.999$ \\
\bottomrule
\end{tabular}
\par\smallskip
\begin{minipage}{\linewidth}\footnotesize
HM = hierarchical model, BS = Birnbaum--Saunders race, WR = Weibull race, and DDM = drift diffusion model. Smaller information criteria indicate better fit after allowing for model complexity. For $k$ estimated parameters and $N=500$ students, the consistent AIC is $\mathrm{CAIC}=\mathrm{BIC}+k$ and the sample-size-adjusted BIC is $\mathrm{SABIC}=-2\ell+k\log[(N+2)/24]$, where $\ell$ is the full-data marginal log likelihood.
\end{minipage}
\end{table}

The hierarchical model predicted withheld responses best, with nearly all of its advantage over logistic diffusion coming from response times. The logistic diffusion analysis showed that items 2 and 3 placed very different time demands on students even at the same predicted chance of success and the same speed. It also revealed a limit of the model in this booklet. Among students whose estimated ability was below item difficulty, later responses were more likely to earn full credit. The model predicted the reverse.

\section{Discussion}
\label{sec:discussion}

In this study, we have shown that the 2PL model can be obtained as the stationary response model of an evidence accumulation process, and it is developed from the same process, a joint model for response accuracy and response time. The evidence accumulation process is a diffusion whose drift moves the accumulated evidence towards $a_i(\theta_p - \beta_i)$ and whose stationary distribution is logistic. Applying a two-boundary rule to the process gives the logistic diffusion IRT model, for which the connection with the P\"oschl--Teller potential yields explicit exit probabilities, first passage time distributions and mean first passage times. In simulation studies, response times improved ability estimates, and the evidence threshold had a larger effect on the accuracy of ability estimates than test length. In PISA 2018 mathematics data, the model distinguished items of similar difficulty by their time demands and predicted withheld answers nearly as accurately as the hierarchical model, while its predictions for response times and for accuracy below item difficulty were less successful.

The model clarifies what response times can and cannot reveal about ability. Response times are informative about the distance between ability and difficulty, while the response indicates its direction. This is consistent with the $12.5\%$ reduction in ability RMSE from joint scoring in Study~1, and why a short response time on PISA item~8 could not distinguish students above from those below the item difficulty. The model also separates two reasons why an item may take long  to answer. An item may require more evidence, reflected in its evidence threshold, or its evidence may accumulate more slowly in observed time, reflected in its time scale. Only the first affects the probability of a correct response. PISA items~2 and~3 had the same success rate and differed four-fold in median time at equal chances of success, a difference the model attributed to the time scale. A model for accuracy alone does not distinguish such items.

\subsection{Practical recommendations}
We have several recommendations for researchers and practitioners who collect response times.

First, response times should be used in item calibration when they are available, and the choice of response time model should be justified. Candidate models can be compared by cross-validated predictive log scores for answers and times separately, so that it clarifies whether the model advantage lies in predicting accuracy or response time.

Second, the evidence threshold and time scale of each item should be reported alongside its difficulty and discrimination. Together they indicate whether an item is slow because it requires much evidence or because its evidence accumulates slowly, which is relevant when tests are assembled under time limits \citep{van2008using}.

Third, test developers should consider how administration conditions affect the amount of evidence respondents require. In Study~3, lowering the evidence threshold from $3.5$ to $1.8$ reduced the accuracy of ability estimates more than halving the number of items. Conditions that encourage responding on less evidence, such as strict time limits or instructions that emphasize speed \citep{voss2004interpreting}, may therefore reduce measurement precision more than a shorter test.

Fourth, response time data should be screened for very short and very long times before calibration. Rapid guessing increased ability RMSE for all models in Study~3, and in the PISA data one group of three withheld responses, including a sub-second time, accounted for about $56\%$ of the difference in predictive performance between the logistic diffusion IRT model and the hierarchical model.

\subsection{Limitations}

The study has several limitations. First, discrimination was underestimated for highly discriminating items, and the bias did not decrease when the sample size was doubled. We attribute this to the numerical approximation used in calibration. Second, the coverage of posterior intervals fell below the nominal level when response times were used in scoring, particularly with higher evidence thresholds and longer tests, because calibration uncertainty was not propagated. Third, the model assumes symmetric boundaries and a starting value of zero, so that correct and incorrect responses are treated symmetrically, and persons equally far above and below an item difficulty have the same response time distribution. This symmetry may be appropriate for many attitude and personality items, but in ability tests, respondents far below an item difficulty may not respond as quickly as those far above it. Fourth, the model describes a single stage of evidence accumulation. The PISA analysis suggests that this is insufficient for certain items, where response time includes the construction of an answer and partial credit was scored as incorrect. Fifth, the model contains no component for disengaged responding or contaminant response times, and it assigned very low density to a sub-second response in the PISA data. 

\subsection{Extensions and future research}

Several extensions follow directly from the structure of the model. Given that the 2PL stationary probability does not depend on the starting value, item-specific starting values or asymmetric boundaries could be introduced to represent response bias without changing the stationary 2PL. Such an extension would relax the symmetry discussed above and allow the probability of a correct response for respondents far below an item difficulty to differ from that for respondents far above it. Person-specific evidence thresholds would allow individual differences in how much evidence respondents require, linking the model to the speed--accuracy trade-off. Mixture components for rapid guessing \citep{wang2015mixture, ulitzsch2020hierarchical} and contaminant response times \citep{ratcliff2002estimating} would address the sensitivity to very short times observed in Study~3 and in the PISA analysis.

The model could also be extended to other response formats and test designs. For constructed-response and polytomous items, several thresholds or several stages of evidence accumulation could represent partial credit and the construction of an answer. Replacing $a_i(\theta_p - \beta_i)$ by a linear combination of several latent variables would give a multidimensional version. Boundaries that change over time could represent time limits and not-reached items, and allowing the evidence state at the end of one item to affect the starting value for the next would provide a process account of item position and context effects. Analogous constructions with other stationary distributions may yield process interpretations of other item response functions.

Methodologically, more accurate numerical approximation of the likelihood is needed to recover highly discriminating items, and Bayesian estimation would allow calibration uncertainty to be propagated to ability estimates. Empirically, the interpretation of the evidence threshold could be tested experimentally. If instructions or time limits that change the emphasis on speed alter the estimated evidence thresholds while leaving difficulty and discrimination unchanged, as validation studies of diffusion model parameters have examined for boundary separation, this would support the distinction the model draws between item parameters and the decision rule. Applications to other assessments, item formats and populations are needed to establish how generally the predicted conditional dependence holds.

\subsection{Conclusion}

The 2PL model has long been used as a description of response accuracy without an account of how responses are produced. This paper shows that the 2PL is the stationary response model of an evidence accumulation process, in which ability, difficulty and discrimination keep their usual meaning. Response time is then modelled by the time the same process takes to reach a decision boundary. Under this account, the accuracy of a response depends on how much evidence a respondent requires as well as on how far ability lies from difficulty, and response times reveal that distance but not its direction. The practical implication is that the conditions under which responses are given, and not only the items, determine how much a test can tell us about ability.

\section*{Statement on AI usage}
GPT-6 Astra was used to proofread this paper, including verification of the mathematical results. It was also used to generate the code for implementing the models in R presented in the paper.
\bibliographystyle{plainnat}
\bibliography{Paper}
\appendix
\section{Appendix}

\subsection{Proof of Theorem \ref{thm:invariant}}\label{app:proof-invariant}

\begin{proof}
Let
\[
        b(x)=-\frac{\sigma^2}{2}\tanh\left(\frac{x-\mu}{2}\right).
\]
Then
\[
        \frac{\mathrm{d} b(x)}{\mathrm{d} x}=-\frac{\sigma^2}{4}\sech^2\left(\frac{x-\mu}{2}\right),
\]
so \(b\) is bounded and globally Lipschitz. The standard existence and pathwise uniqueness theorem for stochastic differential equations gives a unique strong solution for all times.

We now identify the invariant density. A scale density for the diffusion is
\[
        \frac{\mathrm{d} s(x)}{\mathrm{d} x}=\exp\left\{-\int^x \frac{2b(y)}{\sigma^2}\dd y\right\}.
\]
Since
\[
        \frac{2b(y)}{\sigma^2}=-\tanh\left(\frac{y-\mu}{2}\right),
\]
we may choose
\begin{equation}\label{eq:scale-density}
        \frac{\mathrm{d} s(x)}{\mathrm{d} x}=2\cosh^2\left(\frac{x-\mu}{2}\right)
        =1+\cosh(x-\mu).
\end{equation}
Thus one convenient scale function is
\begin{equation}\label{eq:scale}
        S(x)=x-\mu+\sinh(x-\mu).
\end{equation}
The corresponding speed density is
\begin{equation}\label{eq:speed-density}
        \frac{\mathrm{d} m(x)}{\mathrm{d} x}=\frac{2}{\sigma^2\frac{\mathrm{d} s(x)}{\mathrm{d} x}}
        =\frac{1}{\sigma^2}\sech^2\left(\frac{x-\mu}{2}\right).
\end{equation}
The scale function satisfies \(S(x)\to-\infty\) as \(x\to-\infty\) and \(S(x)\to\infty\) as \(x\to\infty\). Also,
\[
        \int_{-\infty}^{\infty}\frac{\mathrm{d} m(x)}{\mathrm{d} x}\dd x
        =\frac{1}{\sigma^2}\int_{-\infty}^{\infty}\sech^2\left(\frac{x-\mu}{2}\right)\dd x
        =\frac{4}{\sigma^2}<\infty.
\]
The scale transform \(S(Z_t)\) is a regular diffusion in natural scale on \(\mathbb R\). It is recurrent and has a finite speed measure, so \(Z_t\) is positive recurrent; see \citet[Theorem V.46.12 and Section V.53]{rogers2000diffusions}.

Normalizing the speed density gives
\[
        \frac{\frac{\mathrm{d} m(x)}{\mathrm{d} x}}{\int \frac{\mathrm{d} m(y)}{\mathrm{d} y}\dd y}
        =\frac14\sech^2\left(\frac{x-\mu}{2}\right)
        =\pi_\mu(x).
\]
This proves that \(\pi_\mu\) is the invariant density. For completeness, we also check the stationary equation directly. The density \(\pi_\mu\) satisfies
\[
        \frac{\sigma^2}{2}\frac{\mathrm{d} \pi_\mu(x)}{\mathrm{d} x}-b(x)\pi_\mu(x)=0,
\]
so the stationary probability current is zero.

The distribution function corresponding to \(\pi_\mu\) is
\[
        F_\mu(x)=\frac{1}{1+\exp\{-(x-\mu)\}}.
\]
Hence
\[
        \int_0^\infty\pi_\mu(z)\dd z
        =1-F_\mu(0)
        =\frac{1}{1+\exp(-\mu)}.
\]
By \citet[Theorem V.54.5]{rogers2000diffusions}, applied to the scale transform \(S(Z_t)\), the transition probabilities of \(Z_t\) converge in total variation to the invariant law with density \(\pi_\mu\). Hence \eqref{eq:logistic-tail} holds.
\end{proof}

\subsection{Proof of Corollary \ref{cor:2pl}}\label{app:proof-2pl}

\begin{proof}
This is Theorem \ref{thm:invariant} with \(\mu=\mu_{ip}\).
\end{proof}

\subsection{Proof of Theorem \ref{thm:laplace-fpt}}\label{app:proof-laplace-fpt}

\begin{proof}
We prove the up-crossing case first. If \(x<c\), the function \(\phi_\lambda\) must solve
\begin{equation}\label{eq:fpt-ode}
        \frac{\sigma^2}{2}\frac{\mathrm{d}^2 \phi(x)}{\mathrm{d} x^2}
        -\frac{\sigma^2}{2}\tanh\left(\frac{x-\mu}{2}\right)\frac{\mathrm{d} \phi(x)}{\mathrm{d} x}
        =\lambda\phi(x),
\end{equation}
with \(\phi(c)=1\) and \(\phi(x)\to0\) as \(x\to-\infty\). We now solve this equation.

Put
\[
        \phi(x)=\cosh(w)\psi(w),
        \qquad w=\frac{x-\mu}{2}.
\]
Since \(\dd w/\dd x=1/2\), direct differentiation gives
\[
        \frac{\mathrm{d} \phi(x)}{\mathrm{d} x}=\frac12\{\sinh(w)\psi(w)+\cosh(w)\frac{\mathrm{d} \psi(w)}{\mathrm{d} w}\},
\]
\[
        \frac{\mathrm{d}^2 \phi(x)}{\mathrm{d} x^2}=\frac14\{\cosh(w)\psi(w)+2\sinh(w)\frac{\mathrm{d} \psi(w)}{\mathrm{d} w}+\cosh(w)\frac{\mathrm{d}^2 \psi(w)}{\mathrm{d} w^2}\}.
\]
Substituting into \eqref{eq:fpt-ode} and simplifying gives
\begin{equation}\label{eq:pt-ode}
        \frac{\mathrm{d}^2 \psi(w)}{\mathrm{d} w^2}+\{2\sech^2w-\nu^2\}\psi(w)=0,
        \qquad
        \nu^2=1+\frac{8\lambda}{\sigma^2}.
\end{equation}
This is the elementary P\"oschl-Teller equation that also underlies Remark \ref{rem:pt}.
Two solutions of \eqref{eq:pt-ode} are
\[
        \psi_+(w)=e^{\nu w}\{\tanh w-\nu\},
        \qquad
        \psi_-(w)=e^{-\nu w}\{\tanh w+\nu\}.
\]
Indeed, differentiating \(\psi_+\) gives
\[
        \frac{\mathrm{d} \psi_+(w)}{\mathrm{d} w}=e^{\nu w}\{\nu(\tanh w-\nu)+\sech^2w\},
\]
\[
        \frac{\mathrm{d}^2 \psi_+(w)}{\mathrm{d} w^2}=e^{\nu w}\{\nu^2(\tanh w-\nu)+2\nu\sech^2w-2\sech^2w\tanh w\}.
\]
Therefore
\[
        \frac{\mathrm{d}^2 \psi_+(w)}{\mathrm{d} w^2}+\{2\sech^2w-\nu^2\}\psi_+(w)=0.
\]
The calculation for \(\psi_-\) is the same. For \(\lambda>0\), \(\nu>1\). As \(w\to-\infty\),
\[
        \cosh(w)\psi_+(w)\to0,
        \qquad
        \cosh(w)\psi_-(w)\to\infty.
\]
Thus the solution satisfying the condition at \(-\infty\) is proportional to \(\cosh(w)\psi_+(w)\). Imposing \(\phi(c)=1\) gives \eqref{eq:up-laplace}.

We now justify that the solution we found is the Laplace transform of the hitting time. Let \(a<x<c\), and let \(\tau_{a,c}=\inf\{t\ge0:Z_t\in\{a,c\}\}\). Apply Ito's formula to \(e^{-\lambda t}\phi(Z_t)\) stopped at \(\tau_{a,c}\). Since \(\phi\) solves \eqref{eq:fpt-ode}, the drift term is zero and hence
\[
        \phi(x)=\E_x\left[e^{-\lambda\tau_{a,c}}\phi(Z_{\tau_{a,c}})\right].
\]
Using the boundary values at \(a\) and \(c\), this becomes
\[
        \phi(x)
        =\E_x[e^{-\lambda\tau_c}\mathbf 1_{\{\tau_c<\tau_a\}}]
        +\phi(a)\E_x[e^{-\lambda\tau_a}\mathbf 1_{\{\tau_a<\tau_c\}}].
\]
Letting \(a\to-\infty\), the second term tends to zero because \(\phi(a)\to0\), and recurrence gives \(\tau_c<\infty\) almost surely. Thus \(\phi(x)=\E_x[e^{-\lambda\tau_c}]\).

For the down-crossing case, reflect the process about its centre. Define \(\widetilde Z_t=2\mu-Z_t\) and \(\widetilde W_t=-W_t\). Since \(-W_t\) is Brownian motion and \(\tanh(-u)=-\tanh(u)\), the process \(\widetilde Z\) solves the same stochastic differential equation as \(Z\), with initial value \(2\mu-x\). The event that \(Z\) hits \(c\) is the same as the event that \(\widetilde Z\) hits \(2\mu-c\). Applying the up-crossing formula to the reflected pair gives \eqref{eq:down-laplace}.
\end{proof}

\subsection{Laplace inversion identities}\label{app:laplace-identities}

The following identities are used to derive the first passage time density in Theorem \ref{thm:fpt-density}.

\begin{lemma}[Two Laplace inversions]\label{lem:laplace-inversions}
Let \(d>0\). For \(s>0\),
\begin{equation}\label{eq:inv-one}
        \mathcal L^{-1}\{e^{-d\sqrt s}\}(t)
        =\frac{d}{2\sqrt\pi\,t^{3/2}}e^{-d^2/(4t)}.
\end{equation}
If \(B\in\R\), then
\begin{align}\label{eq:inv-two}
        \mathcal L^{-1}\left\{\frac{e^{-d\sqrt s}}{\sqrt s-B}\right\}(t)
        &=\frac{1}{\sqrt{\pi t}}e^{-d^2/(4t)}
        +B e^{-dB+B^2t}
        \operatorname{erfc}\left(\frac{d}{2\sqrt t}-B\sqrt t\right),\notag\\
        \mathcal L^{-1}\left\{\frac{e^{-d\sqrt s}}{\sqrt s+B}\right\}(t)
        &=\frac{1}{\sqrt{\pi t}}e^{-d^2/(4t)}
        -B e^{dB+B^2t}
        \operatorname{erfc}\left(\frac{d}{2\sqrt t}+B\sqrt t\right).
\end{align}
\end{lemma}

\begin{proof}
Formula \eqref{eq:inv-one} follows from the integral identity
\[
        \int_0^\infty t^{-3/2}\exp\left(-st-\frac{d^2}{4t}\right)\dd t
        =\frac{2\sqrt\pi}{d}e^{-d\sqrt s}.
\]
For \eqref{eq:inv-two}, first suppose \(\sqrt s>B\). Then
\[
        \frac{e^{-d\sqrt s}}{\sqrt s-B}
        =\int_0^\infty e^{By}e^{-(d+y)\sqrt s}\dd y.
\]
Using \eqref{eq:inv-one} inside the integral gives
\[
        \int_0^\infty e^{By}\frac{d+y}{2\sqrt\pi\,t^{3/2}}
        \exp\left(-\frac{(d+y)^2}{4t}\right)\dd y.
\]
Put \(r=d+y\). This becomes
\[
        \frac{e^{-dB}}{2\sqrt\pi\,t^{3/2}}
        \int_d^\infty r\exp\left(-\frac{r^2}{4t}+Br\right)\dd r.
\]
Completing the square and setting \(u=(r-2Bt)/(2\sqrt t)\) yields
\[
        \frac{1}{\sqrt{\pi t}}e^{-d^2/(4t)}
        +B e^{-dB+B^2t}
        \operatorname{erfc}\left(\frac{d}{2\sqrt t}-B\sqrt t\right).
\]
The identity extends to all real \(B\) by analytic continuation in \(B\). Replacing \(B\) by \(-B\) gives the second formula in \eqref{eq:inv-two}.
\end{proof}

\subsection{Proof of Theorem \ref{thm:fpt-density}}\label{app:proof-fpt-density}

\begin{proof}
We prove the up-crossing formula. Starting from \eqref{eq:up-laplace}, write \(\nu=\sqrt\delta\sqrt{\lambda+1/\delta}\). With the definitions in \eqref{eq:ABC-up},
\[
        e^{-\nu(w_c-w)}=e^{-d\sqrt{\lambda+1/\delta}}
\]
and
\begin{align*}
        \frac{\tanh w-\nu}{\tanh w_c-\nu}
        &=1+\frac{\tanh w_c-\tanh w}{\nu-\tanh w_c} \\
        &=1+\frac{C_+}{\delta}\frac{1}{\sqrt{\lambda+1/\delta}-B}.
\end{align*}
Thus
\[
        \phi_\lambda(x;c)
        =\frac{\cosh w}{\cosh w_c}
        \left[
        e^{-d\sqrt{\lambda+1/\delta}}
        +\frac{C_+}{\delta}
        \frac{e^{-d\sqrt{\lambda+1/\delta}}}{\sqrt{\lambda+1/\delta}-B}
        \right].
\]
Apply Lemma \ref{lem:laplace-inversions} and the shift rule
\[
        \mathcal L^{-1}\{F(\lambda+1/\delta)\}(t)=e^{-t/\delta}f(t),
        \qquad f=\mathcal L^{-1}\{F\}.
\]
This gives \eqref{eq:fpt-density-up}. The down-crossing case follows in the same way from \eqref{eq:down-laplace}, using the second identity in \eqref{eq:inv-two}. If \(x=c\), the hitting time is zero by definition.
\end{proof}

\subsection{Proof of Theorem \ref{thm:mean-fpt}}\label{app:proof-mean-fpt}

\begin{proof}
The mean satisfies
\begin{equation}\label{eq:mean-ode}
        \mathcal L m_c(x)=-1,
        \qquad m_c(c)=0.
\end{equation}
Consider first \(x<c\). Write \(r(x)=\frac{\mathrm{d} m_c(x)}{\mathrm{d} x}\). Then \eqref{eq:mean-ode} becomes
\[
        \frac{\mathrm{d} r(x)}{\mathrm{d} x}-\tanh\left(\frac{x-\mu}{2}\right)r(x)
        =-\frac{2}{\sigma^2}.
\]
The integrating factor is
\[
        I(x)=\sech^2\left(\frac{x-\mu}{2}\right).
\]
Thus
\[
        \frac{\dd}{\dd x}\{r(x)I(x)\}
        =-\frac{2}{\sigma^2}I(x).
\]
To avoid an exponentially growing derivative as \(x\to-\infty\), integrate from \(-\infty\) to \(x\). Since
\[
        \int_{-\infty}^x\sech^2\left(\frac{u-\mu}{2}\right)\dd u
        =2\left\{1+\tanh\left(\frac{x-\mu}{2}\right)\right\},
\]
we get
\[
        r(x)
        =-\frac{4}{\sigma^2}
        \frac{1+\tanh((x-\mu)/2)}{\sech^2((x-\mu)/2)}
        =-\frac{2}{\sigma^2}\{1+e^{x-\mu}\}.
\]
Using \(m_c(c)=0\),
\[
        m_c(x)=-\int_x^c r(u)\dd u
        =\frac{2}{\sigma^2}\left[(c-x)+e^{c-\mu}-e^{x-\mu}\right].
\]

For \(x>c\), integrate the same first-order equation from \(x\) to \(\infty\), again choosing the integration constant so that no exponentially growing term appears. This gives
\[
        r(x)=\frac{2}{\sigma^2}\{1+e^{-(x-\mu)}\}.
\]
Therefore
\[
        m_c(x)=\int_c^x r(u)\dd u
        =\frac{2}{\sigma^2}\left[(x-c)+e^{-(c-\mu)}-e^{-(x-\mu)}\right].
\]
This proves \eqref{eq:mfpt}.
\end{proof}

\subsection{Proof of Theorem \ref{thm:exit-prob}}\label{app:proof-exit-prob}

\begin{proof}
The function \(p_{\ell,r}\) satisfies
\[
        \mathcal L p_{\ell,r}(x)=0,
        \qquad
        p_{\ell,r}(\ell)=0,
        \qquad
        p_{\ell,r}(r)=1.
\]
Writing out the differential equation gives
\[
        \frac{\mathrm{d}^2 p(x)}{\mathrm{d} x^2}-\tanh\left(\frac{x-\mu}{2}\right)\frac{\mathrm{d} p(x)}{\mathrm{d} x}=0.
\]
Let \(q(x)=\frac{\mathrm{d} p(x)}{\mathrm{d} x}\). Then
\[
        \frac{\mathrm{d} q(x)}{\mathrm{d} x}-\tanh\left(\frac{x-\mu}{2}\right)q(x)=0.
\]
Multiplying by \(\sech^2((x-\mu)/2)\) shows that
\[
        q(x)\sech^2\left(\frac{x-\mu}{2}\right)=C
\]
for some constant \(C\). Hence
\[
        \frac{\mathrm{d} p(x)}{\mathrm{d} x}=C\cosh^2\left(\frac{x-\mu}{2}\right).
\]
Since
\[
        2\cosh^2\left(\frac{x-\mu}{2}\right)=1+\cosh(x-\mu)=\frac{\mathrm{d} S(x)}{\mathrm{d} x},
\]
we have \(p(x)=C_1S(x)+C_2\). The two boundary conditions give \eqref{eq:exit-prob}.
\end{proof}

\subsection{Transition density before a response boundary is reached}\label{app:transition-density}

The coordinates and diffusion scale in \eqref{eq:w-bounds} are used below.

The process
\[
        U_t=\frac{Z_t-\mu}{2}
\]
satisfies
\begin{equation}\label{eq:U-sde}
        \dd U_t=-2D\tanh(U_t)\dd t+\sqrt{2D}\dd W_t.
\end{equation}
Let \(q^D(t,u\mid u_0)\) be the transition density of \(U\), killed when it exits \((u_\ell,u_r)\), with respect to the \(u\)-variable.

For a candidate eigenvalue \(\lambda>0\), the squared frequency is \(k^2=\lambda/D-1\), and \(d=u-u_\ell\) is the distance from the left boundary. The solution normalized to have value zero and derivative one at \(u_\ell\) is
\begin{equation}\label{eq:normalized-eigenfunction}
    v_\lambda(u)=\frac{1}{1+k^2}
    \left\{
       (k^2+\tanh u_\ell\tanh u)\frac{\sin(kd)}{k}
       +(\tanh u-\tanh u_\ell)\cos(kd)
    \right\}.
\end{equation}
This expression is even in \(k\) and real for all \(\lambda>0\), including imaginary frequencies \(k\) when \(0<\lambda<D\). At \(\lambda=D\), the continuous value of \(\sin(kd)/k\) is \(d\), so
\[
    v_D(u)=(u-u_\ell)\tanh u_\ell\tanh u+\tanh u-\tanh u_\ell .
\]
For symmetric transformed boundaries \((-b,b)\), \(v_D(b)=2\tanh b (1-b\tanh b)\). Thus zero frequency is an eigenmode exactly when \(b\tanh b=1\).

\begin{theorem}\label{thm:killed}
The eigenvalues \(0<\lambda_1<\lambda_2<\cdots\) are the distinct positive solutions of \(v_\lambda(u_r)=0\), including \(\lambda=D\) whenever \(v_D(u_r)=0\). For each eigenvalue,
\[
    \Psi_n(u)=v_{\lambda_n}(u),\qquad
    N_n=\int_{u_\ell}^{u_r}\Psi_n(v)^2\dd v.
\]
Then, for \(t>0\),
\begin{equation}\label{eq:killed-density}
        q^D(t,u\mid u_0)
        =\frac{\sech u}{\sech u_0}
        \sum_{n\ge1}e^{-\lambda_nt}
        \frac{\Psi_n(u)\Psi_n(u_0)}{N_n}.
\end{equation}
\end{theorem}

\begin{proof}
The density of \(U\) satisfies the forward equation
\begin{equation}\label{eq:forward-U}
        \frac{\partial q^D}{\partial t}
        =D\frac{\partial^2 q^D}{\partial u^2}
        +2D\tanh(u)\frac{\partial q^D}{\partial u}
        +2D\sech^2(u)q^D,
\end{equation}
with boundary conditions \(q^D(t,u_\ell\mid u_0)=q^D(t,u_r\mid u_0)=0\).

Write
\begin{equation}\label{eq:change-r}
        q^D(t,u\mid u_0)=\frac{\sech u}{\sech u_0}R(t,u\mid u_0).
\end{equation}
Since \(\frac{\mathrm{d} (\sech u)}{\mathrm{d} u}=-\sech u\tanh u\),
\[
        \frac{\partial q^D}{\partial u}=\frac{\sech u}{\sech u_0}\{\frac{\partial R}{\partial u}-\tanh(u)R\},
\]
\[
        \frac{\partial^2 q^D}{\partial u^2}
        =\frac{\sech u}{\sech u_0}
        \{\frac{\partial^2 R}{\partial u^2}-2\tanh(u)\frac{\partial R}{\partial u}+(1-2\sech^2u)R\}.
\]
Substitution into \eqref{eq:forward-U} gives
\begin{equation}\label{eq:R-equation}
        \frac{\partial R}{\partial t}
        =D \frac{\partial^2 R}{\partial u^2}-D\{1-2\sech^2u\}R,
\end{equation}
with zero boundary conditions at \(u_\ell\) and \(u_r\).

We now solve \eqref{eq:R-equation} by separation of variables. The spatial equation is
\begin{equation}\label{eq:spatial}
        -D\frac{\mathrm{d}^2 \Psi(u)}{\mathrm{d} u^2}+D\{1-2\sech^2u\}\Psi(u)=\lambda\Psi(u),
        \qquad
        \Psi(u_\ell)=\Psi(u_r)=0.
\end{equation}
This is a regular second-order boundary-value problem on a finite interval. Its eigenvalues are real, its eigenfunctions are orthogonal and complete in \(L^2(u_\ell,u_r)\), and the heat kernel for \eqref{eq:R-equation} is
\begin{equation}\label{eq:R-series}
        R(t,u\mid u_0)
        =\sum_{n\ge1}e^{-\lambda_nt}
        \frac{\Psi_n(u)\Psi_n(u_0)}{N_n}.
\end{equation}
This follows from the usual eigenfunction expansion for regular second-order problems; see \citet[Chapter I, Sections 1.8, 1.9 and 1.13]{titchmarsh1962eigenfunction}.

For each real Dirichlet eigenfunction, integration by parts in \eqref{eq:spatial} gives
\[
    \lambda\int_{u_\ell}^{u_r}\Psi(u)^2\dd u
       =D\int_{u_\ell}^{u_r}\{\frac{\mathrm{d} \Psi(u)}{\mathrm{d} u}+\tanh(u)\Psi(u)\}^2\dd u .
\]
Equality to zero would require \(\Psi(u)=C\sech u\), which satisfies the boundary conditions only for \(C=0\). Thus every eigenvalue is strictly positive.

Direct differentiation of \eqref{eq:normalized-eigenfunction} verifies that
\[
    \frac{\mathrm{d}^2 v_\lambda(u)}{\mathrm{d} u^2}+\{k^2+2\sech^2 u\}v_\lambda(u)=0,\qquad
    v_\lambda(u_\ell)=0,\qquad \left.\frac{\mathrm{d} v_\lambda(u)}{\mathrm{d} u}\right|_{u=u_\ell}=1 .
\]
The same identities hold at \(k=0\) by continuity. Uniqueness for the initial value problem implies that every nonzero eigenfunction with eigenvalue \(\lambda\) is a nonzero multiple of \(v_\lambda\). Hence the right boundary condition \(v_\lambda(u_r)=0\) characterizes the eigenvalues, and \(N_n>0\) for each one. Substitution into \eqref{eq:R-series} and \eqref{eq:change-r} proves \eqref{eq:killed-density}.
\end{proof}

\subsection{Proof of Theorem \ref{thm:joint-exit}}\label{app:proof-joint-exit}

\begin{proof}
The forward equation \eqref{eq:forward-U} can be written as
\[
        \frac{\partial q^D}{\partial t}=-\frac{\partial J}{\partial u},
\]
where
\begin{equation}\label{eq:current}
        J(t,u\mid u_0)
        =-2D\tanh(u)q^D(t,u\mid u_0)-D\frac{\partial q^D(t,u\mid u_0)}{\partial u}.
\end{equation}
The survival probability of the killed process is
\[
        S(t\mid x)=\int_{u_\ell}^{u_r}q^D(t,u\mid u_0)\dd u.
\]
Differentiating and using \(\frac{\partial q^D}{\partial t}=-\frac{\partial J}{\partial u}\),
\[
        \frac{\dd}{\dd t}S(t\mid x)=J(t,u_\ell\mid u_0)-J(t,u_r\mid u_0).
\]
At the absorbing boundaries \(q^D=0\). Hence
\[
        J(t,u_\ell\mid u_0)=-D\left.\frac{\partial q^D(t,u\mid u_0)}{\partial u}\right|_{u=u_\ell},
        \qquad
        J(t,u_r\mid u_0)=-D\left.\frac{\partial q^D(t,u\mid u_0)}{\partial u}\right|_{u=u_r}.
\]
The rate of probability loss through the upper boundary is \(J(t,u_r\mid u_0)\), while the rate of probability loss through the lower boundary is \(-J(t,u_\ell\mid u_0)\). Therefore
\[
        f_{\tau,Y}(t,1\mid x)=-D\left.\frac{\partial q^D(t,u\mid u_0)}{\partial u}\right|_{u=u_r},
        \qquad
        f_{\tau,Y}(t,0\mid x)=D\left.\frac{\partial q^D(t,u\mid u_0)}{\partial u}\right|_{u=u_\ell}.
\]
Now differentiate the series \eqref{eq:killed-density}. At either boundary the terms coming from differentiating \(\sech u\) vanish because \(\Psi_n(u_\ell)=\Psi_n(u_r)=0\). Thus
\[
        \left.\frac{\partial q^D(t,u\mid u_0)}{\partial u}\right|_{u=u_b}
        =\frac{\sech u_b}{\sech u_0}
        \sum_{n\ge1}e^{-\lambda_nt}
        \frac{\left.\frac{\mathrm{d} \Psi_n(u)}{\mathrm{d} u}\right|_{u=u_b}\Psi_n(u_0)}{N_n},
        \qquad u_b\in\{u_\ell,u_r\}.
\]
Substituting this expression into the two flux formulae gives \eqref{eq:exit-density-upper} and \eqref{eq:exit-density-lower}. Summing over the two exit sides gives \eqref{eq:exit-density-total}.
\end{proof}

\end{document}